\documentclass[a4paper]{article}
\usepackage{a4wide}
\usepackage{graphicx,xcolor}
\usepackage{amsmath}
\usepackage{amsthm}
\usepackage{amssymb, latexsym, mathrsfs}
\usepackage{resizegather}
\usepackage{dsfont}
\usepackage{enumerate}
\usepackage{subcaption}
\usepackage{algorithm}
\usepackage{indentfirst}
\usepackage{color}
\usepackage{transparent}
\usepackage{algorithm}
\usepackage{url}
\usepackage{cite}

\usepackage[affil-it]{authblk}
\usepackage[colorlinks=true,citecolor=blue]{hyperref}

\begin{document}

     \title{\textbf{A Distributed Scalable Architecture using $\mathcal{L}_1$ Adaptive Controllers for Primary Voltage Control of DC Microgrids}}
\author[1]{Daniel O'Keeffe\thanks{Research is supported by the Irish Research Council enterprise partnership scheme (Award No. R16920) in collaboration with University College Cork, Ireland and United Technologies Research Centre Ireland Ltd.}\thanks{Email: \tt\small{danielokeeffe@umail.ucc.ie}; Corresponding author}}
\author[2]{Stefano Riverso\thanks{Email: \tt\small RiversS@utrc.ucc.com}}
\author[2]{Laura Albiol-Tendillo\thanks{Email: \tt\small AlbiolL@utrc.ucc.com}}
\author[1,3]{Gordon Lightbody\thanks{Email: \tt\small g.lightbody@ucc.ie}}
\affil[1]{\textnormal{Control \& Intelligent Systems Group, School of Engineering,
University College of Cork, Ireland}}
\affil[2]{\textnormal{United Technologies Research Centre Ireland Ltd, 4th Floor Penrose Business Centre, Cork, Ireland}}
\affil[3]{\textnormal{MaREI-SFI Research Centre, University College Cork, Ireland}}

     \date{\textbf{Technical Report}\\ January, 2018}

     \maketitle

     \begin{abstract}
       This paper proposes a new distributed control architecture for distributed generation units in heterogeneous DC islanded microgrids. Each unit is equipped with state-feedback baseline and augmenting $\mathcal{L}_1$ adaptive voltage controllers at the primary level of the microgrid control hierarchy. Local controller synthesis is scalable as it only requires information about corresponding units, couplings, and at most, the addition of state-predictor measurements of neighbouring controllers. Global asymptotic stability of the microgrid is guaranteed in a plug-and-play fashion by exploiting Lyapunov functions and algebraic Riccati equations. The performance of the proposed architecture is evaluated using a heterogeneous DC islanded microgrid that consists of 6 DC-DC boost converters configured in a radial and meshed topology. The use of $\mathcal{L}_1$ adaptive controllers achieves fast and robust microgrid voltage stability in the presence of plug-and-play operations, topology changes and unknown load changes. Finally, the distributed architecture is tested on a bus-connected islanded-microgrid consisting of linear resistive load and non-linear DC motor.
     
     \end{abstract}

\newpage

\section{Introduction}
Advances in DC-DC power electronics, has led to the promising emergence of DC islanded microgrids (ImGs) \cite{DeDoncker2014}. DC power distribution can avoid inherent issues associated with AC such as harmonic compensation, reactive power and synchronisation; thus improving power quality, efficiency and reliability. Furthermore, the use of DC can reduce the weight of traditional AC power networks by 10 tons/MW \cite{DeDoncker2014}; which is important for electric vehicle and aircraft applications. Recently, DC ImGs have been deployed in low-voltage DC (LVDC) networks such as telecommunication towers, occupied interior spaces, data centres and traction systems \cite{Symanski,Patterson2012,Becker2011,Elsayed2015}. The next wave of DC ImG applications are expected in large-scale residential, commercial and industrial buildings, and aerospace \cite{Diaz2015,Abdelhafez2009, Wheeler2014}.

The proliferation of mGs has led to a growing importance in the development and commercialisation of mG control systems to ensure safe and efficient operation \cite{Olivares2014}. Over recent years, increasing complexities within large-scale systems (LSS) has led to demands for local and scalable algorithms that can coordinate global performance \cite{Anuradha2013}. Thus, decentralised and distributed control architectures have become an attractive alternative to centralised approaches \cite{Riverso2013,Diaz2014,Kumar2015,Wu2015}. Currently, mG control system manufacturers/vendors offer monolithic solutions i.e. a fixed system with limited flexibility and robustness to uncertainty, as outlined in \cite{OKeeffe2017a,Meng2017}. Many control solutions utilise proprietary platforms which require extensive design details on behalf of vendors. Though proprietary designs are standardised and reliable, this approach increases capital and operational expenditure, commissioning time and complexity of the mG design \cite{Katiraei2017}. Additionally, some key features of autonomous mGs include; the economic dispatch of DGUs, the reduction of operating and maintenance costs, and healthy levels of load-servicing and reserve capacity. As a result, there is a growing trend towards mG control solutions that allow plug-and-play (PnP) capabilities to enable system owners the scalability, flexibility and accessibility to implement and maintain controls in LSS. 

PnP control designs, first outlined in \cite{Stoustrup2009}, have successfully been deployed as primary and secondary controllers in the standard hierarchical control structure of AC \cite{Riverso2015,Riverso2017a} and DC ImGs \cite{Tucci2016c,Tucci2017g,Han2017}. Primary controllers are locally responsible for stable power distribution, while secondary controllers coordinate system voltage levels and improve load-sharing accuracy using low-bandwidth communications (LBC).
PnP controllers maintain operation stability when DGUs and loads are reconfigured without requiring \textit{a priori} knowledge. Global asymptotic stability (GAS) is guaranteed by checking the viability of DGU plug-in/out operations through a local optimisation problem using linear matrix inequalities (LMIs). Furthermore, the technique is scalable as local controllers depend only on knowledge of corresponding DGU and line-couplings. Once DGU plug-in/out, neighbouring controllers are required to retune off-line, resulting in limited robustness. Recently, line-independent \cite{Tucci2016e} and robust \cite{Sadabadi2017a} PnP controllers were proposed to overcome this. However, these PnP techniques are computationally extensive, controller gains are required to discontinuously switch after off-line stability checks are performed, and robustness to network uncertainty is limited.

Adaptive control schemes have recently been proposed to extend performance and functionality within heterogeneous and uncertain large-scale mGs. As market deregulation continues and system owners achieve greater flexibility, mGs will become increasingly heterogeneous; consisting of different DGUs/DSUs, topologies, unknown loads, communications and operations. Thus, architectural and network uncertainty will influence the coordination and control of large-scale mGs. 
Fast and robust performance can be guaranteed in the presence of uncertainty and changing dynamics by incorporating adaptive control techniques that utilise Lyapunov stability theory \cite{Slotine1991}. In
\cite{Nasirian2014,Augustine2015,Josep2014,Dragicevic2014,Nasirian2014a}, adaptation is introduced to the primary and secondary controllers. %of cooperating DGUs and DSUs %
However, these strategies are based on premeditated conditions or linear controllers to provide small-signal adjustments to droop coefficients for dynamic performance when achieving system objectives such as voltage restoration and load-sharing. Furthermore, these strategies depend on accurate system models, specific mG topologies, and do not address well-documented adaptive control problems, as outlined in \cite{Anderson2005}. Historically, adaptive control techniques, particularly the model reference adaptive controller (MRAC), faced difficulties with regards to practical implementation. Though adaptive systems can be found throughout nature, ensuring robustness and fast adaptation in the presence of unknown, time-varying dynamics proved difficult to predict and guarantee. Such limitations led to recent advancements in robust-adaptive control, including closed-loop reference model (CRM) and $\mathcal{L}_1$ adaptive control ($\mathcal{L}_1$AC) techniques \cite{Gibson2012,L12010}. 

The CRM adaptive controller uses a closed-loop reference model in order to guarantee stability and improve transient responses within uncertain environments. The first work to perform a comprehensive stability analysis of a DC ImG incorporating adaptive controllers is found in \cite{Vu2017,Vu2017b}. This work also adjusts the droop coefficients using distributed CRM adaptive secondary controllers. Droop control is an inertial controller, and thus adaptation will reactively act on state deviations within the system. Here, the adaptive laws are adapting to uncertainty of the droop coefficients, as opposed to uncertainty concerning the system dynamics. Moreover, rapidly changing dynamics such as PnP operations are not facilitated. Ultimately, \cite{Vu2017,Vu2017b} require the design of adaptive controllers for each global objective i.e. voltage balancing and load-sharing. By performing robust adaptation at the primary level, the number of adaptive controllers is reduced to one, and a more simplified hierarchical control structure, such as the non-droop control coordination layer in \cite{Tucci2017g,Han2017}, can be designed thereafter. 

This paper proposes a new scalable distributed adaptive architecture at the primary control level of DC ImGs. The $\mathcal{L}_1$AC is proposed as it has achieved promising success in practical applications \cite{Michini2009,Gregory2010,Svendsen2012,Li2009,Zhao2014}. Local DC-DC boost power converters are equipped with decentralised state-feedback (DeSSf) baseline controllers and augmented with $\mathcal{L}_1$ adaptive voltage controllers. The rationale for implementing an augmentation approach as opposed to a fully
adaptive one is that in real systems it is common to have baseline controllers designed to provide reference tracking and
disturbance rejection during nominal operation \cite{Michini2009,Gregory2010,Kumaresan2016}. This approach can also facilitate greater flexibility to a system owner as previously discussed.
The distributed architecture is shown to achieve robust voltage control which adheres to IEEE transient and steady-state performance standards of \cite{IEEE2009}. The architecture is evaluated within the context of;
\begin{itemize}
\item Heterogeneous DC ImG consisting of DC-DC boost converters.
\item Arbitrary topologies; radial, meshed and bus-connected.
\item Parametric uncertainty of converter, coupling and unknown load dynamics.
\item Reconfiguration of DGUs through PnP operations and online topology changes due to line faults.
\end{itemize}
This work follows on from our previous work implementing a decentralised $\mathcal{L}_1$AC architecture within DC ImGs \cite{OKeeffe2018e} by guaranteeing GAS in a PnP fashion. Following \cite{OKeeffe2018e}, GAS can be ensured off-line in a centralised fashion using aggregated vector Lyapunov functions. However, this approach leads to conservative designs; controllers require retuning, as shown in \cite{OKeeffe2017a}. Furthermore, determining the correct controllers to retune can prove difficult as the mG size increases in LSS. The distributed architecture instead guarantees GAS by solving local algebraic Riccati equations (AREs), as outlined in \cite{BABANRAO2004,Pagilla2007,OKeeffe2018a}. Consequently, the architecture requires the local $\mathcal{L}_1$ACs to measure the states of each neighbouring $\mathcal{L}_1$AC. Conveniently, this LBC flow has the same topology as the coupling graph, and as a result, the architecture remains scalable.

The paper is organised as follows. In section 2, the DC ImG model is developed and baseline controllers are designed. Sections 3 details the design and GAS analysis of the distributed $\mathcal{L}_1$AC architecture. Finally, section 4 demonstrates the effectiveness of the proposed designs using radial, meshed and bus-connected DC ImG topologies to conduct PnP, topology change and load disturbance tests.

A version of this work has been submitted to IEEE Transactions on Smart Grid.

\section{DC Islanded Microgrid Model}
\begin{figure*}[!htb]    
 \centering
 \includegraphics[width=\textwidth, height=4cm]{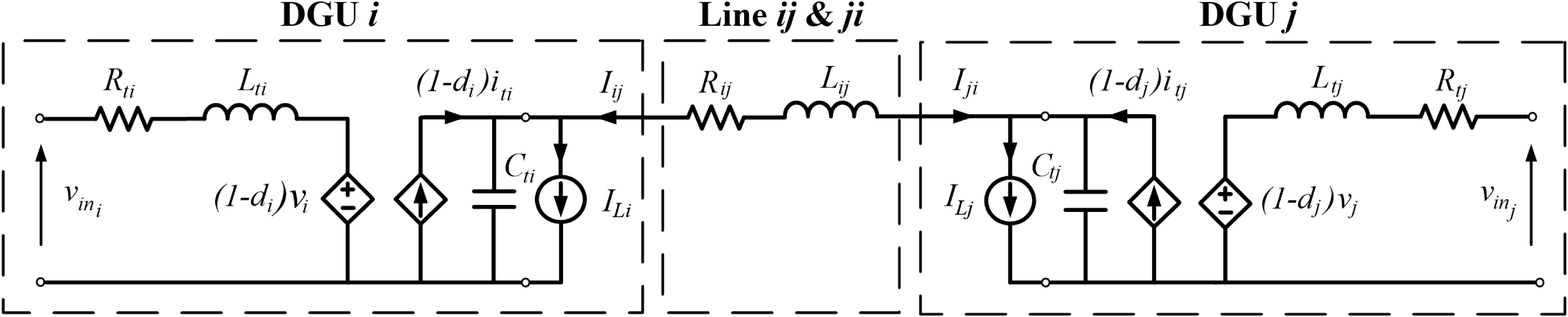}
 \caption{Averaged nonlinear model of DC ImG composed of two coupled boost converter DGUs with unknown loads.}
 \label{fig:MG2}
 \end{figure*}

This work considers boost converters, which step-up low-voltages to high-voltages.  Boost converter controllers are notoriously difficult to tune in mGs due to their non-minimum phase action and have only received attention recently \cite{OKeeffe2017a}, \cite{Sadabadi2017a}.
As proposed in \cite{Tucci2016c}, the DC ImG is modelled as a two-node network for control orientated design. Subsequently, the network is generalised to \textit{N}-nodes. The ImG of Fig. \ref{fig:MG2} is arranged in an arbitrary load-connected topology, where each DGU supplies power to a local load at the point of common coupling (PCC). DGUs can be mapped to load-connections via the Kron Reduction method \cite{D??rfler2013}, which preserves the profile of electrical parameters at the PCC regardless of the topology. This is a positive feature, as the model of each DGU is not dependent on the load, which could be unknown e.g. non-linear/linear resistive, interfacing buck converter or constant power load. Instead, Fig. \ref{fig:MG2} represents the load as a current disturbance, $I_{Li}$. Each DGU is controlled by adjusting the duty-cycle $d_i$ of a solid-state switch using pulse-width modulation (PWM). Fig. \ref{fig:MG2} represents the averaged dynamics of two coupled boost converters, $i$ and $j$, over both on/off switching states. DGUs are coupled via resistive and inductive power lines.

Applying Kirchoff's voltage and current laws to the DC ImG of Fig. \ref{fig:MG2} yields the following set of averaged differential equations:
\begin{subequations}
\begin{equation}
\textrm{DGU $i$:}
\begin{cases}
	     	     \dfrac{d I_{ti}}{dt} = \dfrac{1}{L_{ti}}V_{in_{i}} - \dfrac{(1-d_i)}{L_{ti}} V_{dc_{i}} - \dfrac{R_{ti}}{L_{ti}} I_{ti} \\
	     	     \\
	     \dfrac{d V_{dc_{i}}}{dt} =  \dfrac{(1-d_i)}{C_{ti}} I_{ti} + \dfrac{1}{C_{ti}} I_{ij} - \dfrac{1}{C_{ti}}I_{Li}
	     \end{cases}
	     \label{eq:DGU i}
\end{equation}
\begin{equation}
\textrm{DGU $j$:}
\begin{cases}
	     	     \dfrac{d I_{tj}}{dt} = \dfrac{ 1}{L_{tj}}V_{in_{j}} - \dfrac{(1-d_j)}{L_{tj}} V_{dc_{j}} - \dfrac{R_{tj}}{L_{tj}} I_{tj}, \\
	     	     \\
	     \dfrac{dV_{dc_{j}}}{dt} =  \dfrac{(1-d_j)}{{C_{tj}}} I_{tj} + \dfrac{1}{C_{tj}} I_{ji} - \dfrac{1}{C_{tj}}I_{Lj}
	     \end{cases}
	     \label{eq:DGU j}
\end{equation}
\begin{equation}
\textrm{Line $ij$:}
\begin{cases}
L_{ij}\dfrac{d I_{ij}}{dt} = V_{dc_{j}} - R_{ij} I_{ij} - V_{dc_{i}}.
\end{cases}
\label{eq:LineIJ}
\end{equation}
\begin{equation}
\textrm{Line $ji$:}
\begin{cases}
L_{ji}\dfrac{d I_{ji}}{dt} = V_{dc_{j}} - R_{ji} I_{ji} - V_{dc_{i}}
\end{cases}
\label{eq:LineJI}
\end{equation}
\label{eq:AvMG}
\end{subequations}
\vspace{3mm} 
 \newline
\textbf{Assumption 1:} \textit{
To ensure $I_{ij}(t)$ = $- I_{ji}(t)$ $\forall t \geq 0$, initial line current states are defined as $I_{ij}(0)$ = $- I_{ji}(0)$. With this, $R_{ij}$ = $R_{ji}$ and $L_{ij}$ = $L_{ji}$.}

\subsection{Quasi Stationary Line Model}\label{sec:QSL}
If the time constant of the line transients is very fast, i.e. assuming $L_{ij}$ and $L_{ji}$ are significantly small, then line dynamics can be neglected. This type of model is known as a Quasi-Stationary Line (QSL) approximation. This is usually a good approximation for small-scale mGs where the lines are predominantly resistive. In open-loop, global stability can be inferred by ensuring local DGU stability, as detailed in section 6.1 of \cite{OKeeffe2018e}.
Line equations (\ref{eq:LineIJ}) and (\ref{eq:LineJI}) are represented in steady-state form using QSL approximations, i.e. $\dfrac{d I_{ij}}{dt} = \dfrac{d I_{ji}}{dt}= 0$:

\begin{equation}
I_{ij} = \frac{V_{dc_{j}} - V_{dc_{i}}}{R_{ij}},
\label{eq:i_ij}
\end{equation} 

\begin{equation}
I_{ji} = \frac{V_{dc_{i}} - V_{dc_{j}}}{R_{ji}}.
\label{eq:i_ji}
\end{equation} 

Replacing line current variable $I_{ij}$ of  equation (\ref{eq:DGU i}) 
with equation (\ref{eq:i_ij}) yields the following model for DGU $i$,

\begin{equation}
\textrm{DGU $i$:}
\begin{cases}
	     	     \dfrac{d I_{ti}}{dt} = \dfrac{1}{L_{ti}}V_{in_{i}} - \dfrac{(1-d_i)}{L_{ti}} V_{dc_{i}} - \dfrac{R_{ti}}{L_{ti}} I_{ti} \\
	     	     \\
	     \dfrac{dV_{dc_{i}}}{dt} =  \dfrac{(1-d_i)}{{C_{ti}}} I_{ti} + \dfrac{V_{dc_{j}}}{R_{ij}C_{ti}} - \dfrac{V_{dc_{i}}}{R_{ij}C_{ti}} - \dfrac{1}{C_{ti}}I_{Li}
	     \end{cases}
	     \label{eq:DGU i 2}
\end{equation}

Interchanging indexes $i$ and $j$ yields the model for DGU $j$. Representing (\ref{eq:DGU i 2}) in a general compact state space form, the dynamics of DGU $i$ are,

\begin{equation}
\Sigma_{[i]}^{\textrm{DGU}}:
\begin{cases}
\dot{x}_{[i]}(t) = \left[ \begin{array}{cc}
-\frac{R_{ti}}{L_{ti}} & -\frac{(1-d_i)}{L_{ti}}\\
\frac{(1-d_j)}{C_{tj}} & -\frac{1}{R_{ij}C_{ti}}
\end{array} \right]x_{[i]}(t)+\left[ \begin{array}{c}
\frac{1}{L_{t_{i}}}\\
0
\end{array} \right]V_{in_{i}} + \left[ \begin{array}{c}
0\\
-\frac{1}{C_{ti}}
\end{array} \right]I_{Li} + \left[ \begin{array}{cc}
0 & 0 \\
0 & \frac{1}{R_{ij}C_{ti}}
\end{array} \right]x_{[j]}(t)
\\
y_{[i]}(t) = C_ix_{[i]}(t)
\end{cases}
\label{eq:DGUSS}
\end{equation}
where ${x}_{[i]}(t)= [I_{t_{i}}, V_{dc_{i}}]^T$ , $I_{Li}$ is the exogenous current disturbance. Unlike with the buck converter, where the averaged state space model of (\ref{eq:DGUSS}) is equivalent to the small-signal state space model, the boost converter is different. From the state matrix of above, the duty-cycle control input is a product of the state vector. As a result, the duty-cycle operating point directly influences stability. The averaged model is therefore non-linear and must be linearised about the duty-cycle operating point by forming a small-signal model\footnote{Note: each average quantity can be expressed as the sum of its steady state and small-signal values e.g. $d_k = D_k + \tilde d_k$, $V_{dc_{k}} = \bar{V}_{dc_{k}} + \tilde v_{dc_{k}}$.}.

\begin{equation}
\Sigma_{[i]}^{\textrm{DGU}}:
\begin{cases}
\dot{x}_{[i]}(t) = A_{ii}x_{[i]}(t)+B_{i}u_{[i]}(t) + E_{i}d_{[i]}(t) + \zeta_{[i]}(t) + \gamma_{[i]}(t) \\
y_{[i]}(t) = C_ix_{[i]}(t)
\end{cases}
\label{eq:DGUSS2}
\end{equation}
where ${x}_{[i]}(t)= [\tilde i_{t_{i}},\tilde v_{dc_{i}}]^T$ , is the small-signal state vector, $u_{[i]}(t) = \tilde{d}_i(t)$ is the small-signal PWM control signal, $d_{i}(t) = \tilde{i}_{Li}$ is the small-signal exogenous current disturbance, $\zeta_{[i]}(t) = A_{ij}x_{j}(t)$ represents coupling with DGU $j$ and $\gamma_i(t) = \frac{\tilde v_{in_{i}}}{L_{ti}}$ is the small-signal input voltage disturbance. It is assumed that changes in input voltages $V_{in_{k}}$ are very slow, and thus can be neglected\footnote{As the input voltage to power converters in a mG is usually from renewable power or storage devices. The dynamics of these devices are much slower than the fast switching dynamics of power converters, therefore it is a safe assumption to neglect small-signal changes in input voltage}. Therefore $\gamma_i(t) = 0$.

The matrices of (\ref{eq:DGUSS2}) are,
\begin{equation*}
A_{ii}=
\left[ \begin{array}{cc}
-\frac{R_{ti}}{L_{ti}} & -\frac{(1-D_i)}{L_{ti}}\\
\frac{(1-D_j)}{C_{tj}} & -\frac{1}{R_{ij}C_{ti}}
\end{array} \right]
A_{ij}=\left[ \begin{array}{cc}
0 & 0 \\
0 & \frac{1}{R_{ij}C_{ti}}
\end{array} \right]
B_{i} = \left[ \begin{array}{c}
\frac{\bar{V}_{dc_{i}}}{L_{t_{i}}}\\
\frac{-\bar{I}_{t_{i}}}{C_{ti}}
\end{array} \right]
E_i = 
\left[ \begin{array}{c}
0\\
-\frac{1}{C_{ti}}
\end{array} \right]
C_i =
\left[ \begin{array}{cc}
0 & 1
\end{array} \right]
\end{equation*}

where $\bar{V}_{dc_{i}} = \frac{\bar{V}_{in_{i}}}{(1-D_i)}$ and $\bar{I}_{t_{i}} = \frac{\bar{V}_{in_{i}}}{(1-D_i)^2R_{L_{i}}}$.

\subsection{QSL Model DC Islanded Microgrid Composed of \textit{N} DGUs}
In this section, the two DGU network of Fig. \ref{fig:MG2} is generalised to an ImG composed of \textit{N} converter DGUs. \cite{OKeeffe2017a} demonstrated that converter coupling dynamics predominantly manifest from physical power lines; duty-cycle  coupling is weak. Neighbouring DGUs are thus defined if they are coupled by the $RL$ power line of Fig. \ref{fig:MG2}. Letting $\mathcal{D} = \{1,...,\textit{N-1}\}$, $\mathcal{N}_i \subset \mathcal{D}$ denotes a neighbour-subset for DGU $i$. As before, assuming QSL approximation of all line dynamics $(i, j)\in \mathcal{D}$, the DC ImG model is represented by (\ref{eq:DGUSS}), with $\zeta_{[i]}(t) = \sum_{j\in\mathcal{N}_i}A_{ij}x_{[j]}(t)$. The only change in (\ref{eq:DGUSS}) is the local state vector matrix $A_{ii}$, becoming:
 \begin{equation}
A_{ii}=
\left[ \begin{array}{cc}
-\frac{R_{ti}}{L_{ti}} & -\frac{(1-D_i)}{L_{ti}}\\
\frac{(1-D_i)}{C_{ti}} & \sum_{j\in\mathcal{N}_i}-\frac{1}{R_{ij}C_{ti}}
\end{array} \right]
\label{eqn:Aii}
 \end{equation}
The overall global model of the $N$ DGU ImG can be given by,
 \begin{equation}
 \mathbf{\Sigma}^{DGU}_{[N]}:
 \begin{cases}
 \dot{\textbf{x}}(t)= \textbf{Ax}(t)+\textbf{Bu}(t) + \textbf{Ed}(t) \\
  \textbf{y}(t) = \textbf{Cx}(t)
  \end{cases}
  \label{eqn:LSSMIMO}
 \end{equation}
where $\textbf{x} = (x_{[1]}, x_{[2]},....,x_{[N]}) \in \mathbb{R}^{2N}, \textbf{u} = (u_{[1]}, u_{[2]},....,u_{[N]}) \in \mathbb{R}^{N}, \textbf{d} = (d_{[1]}, d_{[2]},....,d_{[N]})\in \mathbb{R}^{N}, \textbf{y} = (y_{[1]}, y_{[2]},....,y_{[N]})\in \mathbb{R}^{N}$. Matrices $\textbf{A}$, $\textbf{B}$, $\textbf{C}$ and $\textbf{E}$ are detailed in the section 6.1 of \cite{OKeeffe2018e}.

\subsection{Decentralised Baseline Voltage Control} 
DeSSf baseline controllers are designed for standalone decoupled converters, assuming a connection to a linear resistive load. As in \cite{OKeeffe2018e}, the baseline controllers are designed for decoupled DGUs using \textit{a priori} knowledge of nominal parameters. In order to track constant voltage references in the presence of constant current disturbances, an integral state error between the reference voltage and output voltage is added to the local DGU model. The dynamics are defined as,
   \begin{equation}
   \xi_{[i]}(t) = \int_{0}^{t}(V_{ref_{[i]}} - y_{[i]}(t)) dt = \int_{0}^{t}(V_{ref_{[i]}} - C_{i}x_{[i]}(t)) dt
   \end{equation}
   The DeSSf control law with integral action becomes,
  \begin{equation}
  \mathcal{C}_{[i]} : u_{[i]}^{bl}(t) = -K_{i}^{bl}\hat{x}_{[i]}(t)
  \label{eq:SFBL}
  \end{equation}
  where $K_{i}^{bl} = [K_{i}^i, K_{i}^v, K_{i}^\xi] \in \mathbb{R}^{3}$ is the DeSSf control gain vector. Subsequently, the open-loop model augmented with the integral state $\xi_{[i]}(t)$ becomes third order, hence $\bar{x}_{[i]}(t) = [[x_{[i]}(t)]^T, \xi_{[i]}(t)]^T \in \mathbb{R}^{3}$ is the augmented open-loop state vector. The state-space model of $\Sigma_{[i]}^{\textrm{DGU}}$ can now be defined as,
  \begin{equation}
   \hat{\Sigma}_{[i]}^{\textrm{DGU}}:
   \begin{cases}
   \dot{\bar{x}}_{[i]}(t) = \hat{A}_{ii}\bar{x}_{[i]}(t)+ hat{B}_{i}u_{[i]}^{bl}(t) + \bar{E}_{i}\bar{d}_{[i]}(t) + \bar{\zeta}_{[i]}(t) \\
   \bar{y}_{[i]}(t) = \bar{C}_{i}\bar{x}_{[i]}(t)
   \end{cases}
   \label{eq:DGUSSCL}
   \end{equation}
   where $\bar{d}_{[i]} = [d_{[i]}, V_{ref_{[i]}}]^T \in \mathbb{R}^2$ is the exogenous signal vector, which includes load current disturbance and reference voltage, $\bar{\zeta}_{[i]}(t) = \sum_{j\in\mathcal{N}_i}\hat{A}_{ij}\bar{x}_{[j]}(t)$, and $\bar{y}_{[i]}(t)$ is the measurable output. $\bar{A}_{ii} \in \mathbb{R}^{3\textrm{x}3}$, $\hat{B}_{i}\in \mathbb{R}^{3\textrm{x}1}$, $\bar{E}_i \in \mathbb{R}^{3\textrm{x}3}$,  $\bar{A}_{ij} \in \mathbb{R}^{3\textrm{x}3}$ and $\bar{C_i} \in \mathbb{R}^{3}$.  $(\hat{A}_{ii}, \hat{B}_{i})$ is assumed to be controllable, as demonstrated in \cite{Tucci2016c}. Similarly, the matrices of (\ref{eq:DGUSSCL}) are defined in \cite{OKeeffe2018e}. The DeSSf controllers can be tuned via pole placement or using linear quadratic integral (LQI) regulation.

\newpage

\section{Distributed $\mathcal{L}_1$ Adaptive Control Architecture}
\begin{figure}[!htb]    
\centering
\includegraphics[width=8.5cm]{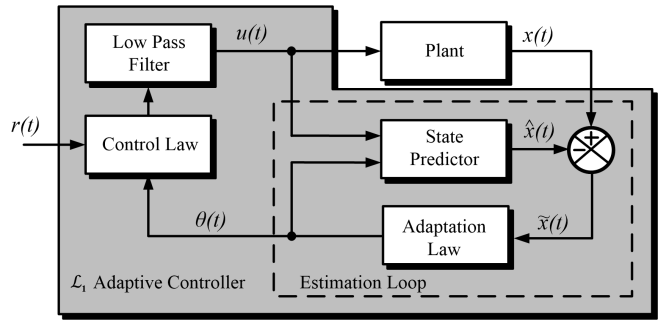}
\caption{General Architecture of $\mathcal{L}_1$ Adaptive Controller. Adapted from \cite{L12010}}.
\label{fig:L1Arch}
\end{figure}

The $\mathcal{L}_1$AC is a modification of the indirect MRAC architecture and was developed to address the issues of providing transient guarantees and determining an optimal rate of adaptation without sacrificing robustness to uncertainty \cite{L12010}. Conventional MRAC suffers from a trade-off between estimation and robustness; large adaptive gains induce high gain feedback which usually leads to high-frequency oscillations in the control-channel that can destabilise the control-loop. The $\mathcal{L}_1$AC architecture decouples the trade-off between estimation and robustness by inserting a low-pass filter (LPF) at the input to both the plant and state-predictor, as seen in Fig. \ref{fig:L1Arch}. Consequently, robustness instead depends on the choice of filter-bandwidth rather than the adaptive gain, thus allowing fast adaptation. Performance bounds and asymptotic stability are derived using Lyapunov based methods.
The $\mathcal{L}_1$AC has achieved successful implementations in various safety-critical applications; notably in aircraft auto-pilots, where the $\mathcal{L}_1$AC is used to maintain stability and good performance in conditions of high aerodynamic uncertainty, and during faults such as, component failure and communication latencies \cite{Gregory2010}. Other $\mathcal{L}_1$AC applications include unmanned water \cite{Svendsen2012} and aerial vehicles \cite{Michini2009}. These applications require controllers that can maintain safe and reliable operation across large operating ranges that experience uncertain conditions. The $\mathcal{L}_1$AC achieves scalable behaviour across different speed ranges, operating dynamics (e.g. surge/cruise), payload and vehicle sizes. Furthermore, $\mathcal{L}_1$AC has been used to augment an LQR baseline controller to enhance the performance tracking enhancement of a decentralised leader-follower coordination scheme for unmanned aerial vehicles \cite{Kumaresan2016}. While receiving great attention in aerospace applications, there are many attractive opportunities for the $\mathcal{L}_1$AC architecture in smart-grid applications. Recently, the $\mathcal{L}_1$AC has been implemented to achieve robust maximum power-point tracking of a wind turbine during uncertainty conditions such as shadow effects, wind shear and speed variations \cite{Zhao2014}.
Ultimately, the $\mathcal{L}_1$AC architecture has potential to improve mG voltage control in heterogeneous and uncertain environments.

From Fig. \ref{fig:L1Arch} a state-predictor replaces the reference model of the conventional MRAC, and a LPF limits the control signal bandwidth. The state-error dynamics, $\tilde x(t)$, between the plant and state-predictor drives the adaptation law. This adjusts the control parameters in order to ensure bounded and asymptotic convergence of state and estimation signals.
 
 \subsection{Plant structure}
The plant has a known structure, but with unknown parameter values.
\vspace{3mm} \newline
\textbf{Assumption 2:} \textit{The design of the distributed $\mathcal{L}_1$AC architecture can neglect the exogenous load disturbance signal ${d}_{[i]}(t)$ as it is compensated by the integral action of $u_{[i]}^{bl}(t)$.}
\vspace{3mm} \newline
 A matched uncertainty term is introduced to represent parametric uncertainty in the dynamics of $\hat{\Sigma}_{[i]}^{\textrm{DGU}}$, hence (\ref{eq:DGUSSCL}) can be represented as,
\begin{equation}
 \hat{\Sigma}_{[i]}^{\textrm{DGU}}:
 \begin{cases}
  \dot{\bar{x}}_{[i]}(t) = \hat{A}_{ii}\bar{x}_{[i]}(t)+ \hat{B}_{i}(u_{[i]}(t) +\theta_{[i]}(t)\hat{x}_{[i]}(t)) + F\bar{E}_i\bar{d}_{[i]}(t)+\bar{\zeta}_{[i]}(t)  \\
 \bar{y}_{[i]}(t) = \bar{C}_i\bar{x}_{[i]}(t)
 \end{cases}
 \label{eq:L1SS}
 \end{equation}
where $\bar{x}_{[i]}(t) \in \mathbb{R}^{3}$, is the system measurable state vector; $u(t) \in  \mathbb{R}$ is the control signal; $F = [0, 0, 1]$; $\theta_{[i]}(t)$ is the unknown matched parametric uncertainty vector. This belongs to a known compact convex set of uniform boundedness $\theta \in \Theta \subset \mathbb{R}^3$.

\subsection{Control Law}
The small-signal control input $u(t)$ for $\hat{\Sigma}_{[i]}^{\textrm{DGU}}$ consists of the summation between the baseline and $\mathcal{L}_1$AC control signals,
\begin{equation}
\mathcal{C}_{[i]}^{\mathcal{L}_1} : u_{[i]}(t) = u_{[i]}^{bl}(t) + u_{[i]}^{\mathcal{L}_1}(t)
\label{eqn:ctrl}
\end{equation}
The Laplace domain representation of the augmenting $\mathcal{L}_1$AC law, fitted with a first-order LPF, is 
\begin{equation}
u_{[i]}^{\mathcal{L}_1}(s) = -C(s)[\hat{\theta}_{[i]}\hat{x}_{[i]}](t)
\label{eqn:C_L1}
\end{equation}
where $C(s)=\frac{\omega_c}{s+\omega_c}$, $\hat{\theta} \in \mathbb{R}^{3}$ is the parametric estimation vector. The robustness of the $\mathcal{L}_1$AC is dependent on the LPF bandwidth $\omega_c$, as subsequently designed.

\subsection{State-Predictor}

\textbf{Remark 1.} \textit{Following section 3.1.3 of \cite{OKeeffe2018e}, as the parameters of $\bar{B}_i$ are unknown, the state-predictor dynamics are defined in control canonical form. For readability, we keep state notation in terms of $x_{[i]}(t)$ as opposed to $z_{[i]}(t)$ in \cite{OKeeffe2018e}.}
\vspace{3mm} \newline
The state-predictor generates an estimate of the system states. From the perspective of the $\mathcal{L}_1$AC, the baseline dynamics are combined with the open-loop DGU dynamics to form an augmented closed-loop system. As the design of the $\mathcal{L}_1$AC is distributed, the structure of the state-predictor requires measurement of neighbouring predictor states. The reasoning becomes apparent when guaranteeing GAS. Without loss of generality, the state-predictor formulation and the desired closed-loop dynamics are equal for all DGUs,
\begin{equation}
 \mathcal{E}_{[i]}:
 \begin{cases}
 \dot{\hat{x}}_{[i]}(t) = \hat{A}_{m}\hat{x}_{[i]}(t)+b(u_{[i]}^{\mathcal{L}_1}(t)+\hat{\theta}_{[i]}(t)\hat{x}_{[i]}(t)) + F\hat{E}_i\hat{d}_{[i]}(t)+\hat{\zeta}_{[i]}(t)  \\
 \hat{y}_{[i]}(t) = \hat{C}_{i}\hat{x}_{[i]}(t)
 \end{cases}
 \label{eq:L1SPSS2}
 \end{equation}
where $\hat{A}_{m} \in \mathbb{R}^{3\textrm{x}3}$ is the Hurwitz design matrix that specifies the desired closed-loop dynamics and $b$ is the input vector in control canonical form of \cite{OKeeffe2018e}. 
 
 \subsection{Local Adaptive Law and Global Asymptotic Stability}
The adaptive law of local $\mathcal{L}_1$ACs generates an estimate of the uncertainties that each DGU experiences, such as unknown parameters,  change in dynamics due to PnP operations,  topology change, unexpected disturbances and possible faults. Ensuring stable adaptation and bounded signals is based on Lyapunov's second method of stability. By defining the Lyapunov function in terms of local state-error and parametric estimation error vectors, the energy trajectories of local system states and estimates remain bounded and local asymptotic stability is guaranteed. Moreover, this paper structures the Lyapunov function as an ARE, the solution of which guarantees GAS, as in \cite{BABANRAO2004,Pagilla2007,OKeeffe2018a}. Local state error is defined as, $\tilde{{x}}_{[i]}(t) = \bar{x}_{[i]}(t) - \hat{x}_{[i]}(t) \in \mathbb{R}^3$ in control canonical form. The estimate error is defined as $\tilde{{\theta}}_{[i]}(t) = \bar{\theta}_{[i]}(t) - \hat{\theta}_{[i]}(t) \in \mathbb{R}^3$. The state-error dynamics, used to drive the adaptive law, is defined as,
\begin{equation}
   \dot{\tilde{x}}_{[i]}(t) = \hat{A}_{m}\tilde{x}_{[i]}(t)+b\tilde{\theta}(t)\bar{x}_{[i]}(t)+\sum_{j\in\mathcal{N}_i}\hat{A}_{ij}\tilde{x}_{[j]}(t)
   \label{eqn:ErrorDyn2}
   \end{equation}
  The quadratic Lyapunov function candidate that describes the aggregated global energy within the system is defined as,
\begin{equation}
\mathcal{V}(\tilde{x}(t), \tilde{\theta}(t)) = \sum\limits_{i = 0}^{N}( \tilde{x}_{[i]}(t)^TP_{i}\tilde{x}_{[i]}(t) + \tilde{\theta}_{[i]}(t)^T\Gamma_{i}^{-1}\tilde{\theta}_{[i]}(t))
\label{eqn:LyapVi}
\end{equation}
where, $P_{i} \in \mathbb{R}^{3\textrm{x}3}$ is a symmetric matrix, such that $P_{i} = P_{i}^T > 0$ is the solution to the Lyapunov linear inequality $\hat{A}_{m}^TP_{i} + P_{i}\hat{A}_{m} \leq -Q_{i}$, for arbitrary $Q_{i} = Q_{i}^T > 0$, and $\Gamma_{i} \in \mathbb{R}^+$ is the adaptive gain.
The derivative of (\ref{eqn:LyapVi}) can be written as,
   \begin{equation}
   \begin{aligned}
   \dot{\mathcal{V}}(t)\leq \sum\limits_{i = 0}^{N}(2(\hat{A}_{m}\tilde{x}_{[i]}+b\tilde{\theta}_{[i]}\hat{x}_{[i]}+\sum_{j\in\mathcal{N}_i}A_{ij}\tilde{x}_{[j]})P_{i}\tilde{x}_{[i]} + \tilde{\theta}_{[i]}\Gamma_{i}^{-1}\dot{\tilde{\theta}}_{[i]})
   \label{Vi2_0}
   \end{aligned}
   \end{equation}
   The adaptive law for the uncertainty estimate is given as,
            \begin{equation}
            \dot{\hat{\theta}}_{[i]} = \Gamma_{i}Proj(\hat{\theta}_{[i]}, -x_{[i]}(t)\tilde{x}_{[i]}^T(t)P_{i}b)
            \label{eqn:adaptiveLaw2i}
            \end{equation}
            The projection operator, described in \cite{L12010}, is used to prevent parametric drift by  upper-bounding the parameter estimate \textit{a priori} i.e. $\theta_{max}$. This, along with the LPF, allows for robust-adaptation. With this, (\ref{Vi2_0}) becomes,
   \begin{equation}
      \begin{aligned}
      \dot{\mathcal{V}}(t) =\sum\limits_{i \leq 0}^{N}(\tilde{x}_{[i]}^T(\hat{A}_{m}^TP_{i} +P_{i}\hat{A}_{m})\tilde{x}_{[i]} +  P_{i}\tilde{x}_{[i]}^T\sum_{j\in\mathcal{N}_i}A_{ij}\tilde{x}_{[j]}+(\sum_{j\in\mathcal{N}_i}A_{ij}\tilde{x}_{[j]})P_{i}\tilde{x}_{[j]})
      \label{Vi2}
      \end{aligned}
      \end{equation}
      Expanding the two summation terms of (\ref{Vi2}) and using the inequality $X^TY + Y^TX \leq X^TX + Y^TY$ from \cite{BABANRAO2004} yields,
                     \begin{equation}
                   \begin{aligned}
                   \dot{\mathcal{V}}_{[i]}(t) \le \sum\limits_{i = 0}^{N}( \tilde{x}_{[i]}^T(\hat{A}_{m}^TP_{i} +P_{i}\hat{A}_{m})\tilde{x}_{[i]} + N_i\tilde{x}_{[i]}^T(P_{i}^TP_{i})\tilde{x}_{[i]}+
                   \sum_{j\in\mathcal{N}_i}\tilde{x}_{[j]}^T(A_{ij}^TA_{ij})\tilde{x}_{[j]})
                   \label{Vidot3}
                   \end{aligned}
                   \end{equation} 
                   \textbf{Assumption 3:} \textit{For conservativeness, knowledge of the upper-bound on the coupling gain matrix $\hat{A}_{ij}$ is assumed for all DGUs.}
                   \vspace{3mm} \newline
                    The coupling term in (\ref{Vidot3}) can be upper bounded as,
                   \begin{equation}
                   \sum_{j\in\mathcal{N}_i}\tilde{x}_{[j]}^T(\hat{A}_{ij}^T\hat{A}_{ij})\tilde{x}_{[j]} \le \Xi_{i}^2\sum_{j\in\mathcal{N}_i}\tilde{x}_{[j]}^T\tilde{x}_{[j]}
                   \end{equation}
                  where $\Xi_{i}^2 \triangleq \sum_{j\in\mathcal{N}_i}\lambda_{max}(\hat{A}_{ij}^T\hat{A}_{ij})=\sum_{j\in\mathcal{N}_i}\lambda_{max}(\hat{A}_{ij}^2)\leq N_i\lambda_{max}(\hat{A}_{ij}^2)$, and $\lambda_{max}$ corresponds to the maximum eigenvalue. Finally, following some index manipulation in \cite{OKeeffe2018e},
                  \begin{equation}
                  \begin{aligned}
                  \dot{\mathcal{V}}(t) 
                  \leq \sum\limits_{i = 0}^{N}\tilde{x}_{[i]}^T(A_{m}^TP_{i}+P_{i}A_{m}+N_iP_{i}P_{i}+\Xi_{i}^2\mathbb{I})\tilde{x}_{[i]}
                  \end{aligned}
                  \label{eq:Vdot5}
                  \end{equation}
                 The global vector Lyapunov function is structured as an ARE such that if there exists a positive-definite matrix $P_{i}$ that solves each local ARE  $A_{m}^TP_{i}+P_{i}A_{m}+N_iP_{i}P_{i}+(\Xi_{i}^2+\epsilon_i)\mathbb{I} = 0$, where $\epsilon_i>0$, 
                 then
                \begin{equation}
                \dot{\mathcal{V}}(t)\leq -\sum\limits_{i = 0}^{N}\tilde{x}_{[i]}^T\varepsilon_{[i]}\tilde{x}_{[i]}
                \label{eqn:Vdotfinal}
                \end{equation}
                To ensure a positive-definite $P_i$ exists for each $\mathcal{L}_1$AC, Lemma's 1 and 2 from \cite{OKeeffe2018a} are invoked. Lemma 1 states that the Hamiltonian matrix associated with the ARE must be hyperbolic for $P_i > 0$ to exist. Lemma 2 states that for such a Hamiltonian matrix to be hyperbolic, the following condition must be satisfied,
                \begin{equation}
               \gamma \triangleq \min\limits_{\omega \in R^+}\sigma_{min}(A_{m}-j\omega \mathbb{I}) > \sqrt{N_i\Xi_i^2}>0
               \label{eqn:distance}
                \end{equation}
                Therefore, the desired closed-loop dynamics should be designed such that the distance between desired eigenvalues and the imaginary axis is greater than $\sqrt{N_i\Xi_i^2}$. To compute the distance $\gamma$, the bisection method of \cite{Aboky2002}%Byers%,
                  , shown in section of 7.5 of \cite{OKeeffe2018a}, is used. Furthermore, Barbalat's Lemma can be invoked to show that global system states converge to the desired states i.e. $\lim_{t\rightarrow \infty}\tilde{x}_{[i]}(t) = 0$.
                 Ultimately, GAS of the overall adaptive DC ImG is guaranteed when the adaptive law is chosen as in (\ref{eqn:adaptiveLaw2i}) and the design condition (\ref{eqn:distance}) is satisfied.
                
                  \subsection{Filter Design}
              The key feature of the $\mathcal{L}_1$AC architecture is the design of a LPF which decouples robustness from adaptation. At this point, GAS has been guaranteed during nominal operation and adaptation. Here, boundedness and stability is further guaranteed when the LPF is inserted. The LPF bandwidth is tuned using the $\mathcal{L}_1$ norm condition. A  reference system is defined for the predictor in order to facilitate performance specification in the Laplace domain. $\hat{\theta}_{[i]}(t) \rightarrow \theta_{[i]}$ is assumed.
              \begin{equation}
              \begin{aligned}
              \hat{x}_{ref_{[i]}}(s) =  (s\mathbb{I}-\hat{A}_{m})^{-1}b((1-C(s))\theta_{[i]}(\hat{x}_{ref_{[i]}}(t)+\tilde{x}_{[i]}(t))+ F\hat{d}_{[i]}(s)+ \hat{\zeta}_{[i]}(s)+\hat{x}_{ic_{[i]}}(s)
              \label{eqn:clref1}
             \end{aligned}
              \end{equation}
              where $\hat{x}_{ref_{[i]}} \in \mathbb{R}^3$, is the reference state vector, $\hat{x}_{ic_{[i]}} \in \mathbb{R}^3$ is the initial state vector, $\mathbb{I} \in \mathbb{R}^{3\textrm{x}3}$, is the identity matrix. From (\ref{eqn:Vdotfinal}), $\tilde{x}_{[i]}(t)$ is bounded, and $\theta_{[i]}$ is bounded by $\theta_{max}$, which represents the boundary of adaptation. Initial states can be assumed bounded, while the reference signal vector $F\hat{d}_{[t]}(t)$ is a constant. To ensure the local reference states remain bounded when the LPF is inserted, a sufficient stability condition via the small-gain theorem is,
                \begin{equation}
                            \lambda = ||G(s)||_{\mathcal{L}_1}\theta_{max} < 1
                            \label{eqn:L1normCond}
                            \end{equation}where, the desired local closed-loop behaviour is represented by the transfer function, $H(s) = (s\mathbb{I}-\hat{A}_{m})^{-1}b$ and $G(s)=H(s)(1-C(s))$. 
               The adaptation bound is defined as,
              \begin{equation}
              \theta_{max} = \max_{\theta \in \Theta}||\theta||_1
              \label{eqn:ThetaMax}
              \end{equation} 
              The states of the overall reference system are bounded if the interconnection term $\hat{\zeta}_{[i]}(t)$ remains bounded. The interconnection term,  $\hat{\zeta}_{[i]}(t)\leq N_i\lambda_{max}(\hat{A}_{ij})x_{ref_{[j]}}(s)$, is bounded if (\ref{eqn:L1normCond}) is also satisfied for $\mathcal{E}_{[j]}$ since condition (\ref{eqn:L1normCond}) applies for all DGUs.
 \subsection{Algorithm for Controller Design}
   Algorithm 1 collects the steps of the overall design procedure.

  \begin{algorithm}
                \caption{Design of distributed controllers $\mathcal{C}_{[i]}^{\mathcal{L}_{1}}$ for subsystem $\hat{\Sigma}_{[i]}^{\textrm{DGU}}$}
              
                \textbf{Input:}  $\mathcal{E}_{[i]}$ as in (\ref{eq:L1SPSS2}), $\mathcal{N}_i$, $\Xi_i^2$, $\theta_{max}$ 
                \\
                \textbf{Output:} Controller $\mathcal{C}_{[i]}^{\mathcal{L}_{1}}$ as in (\ref{eqn:ctrl})
                \newline
                \\ 
                (I) Design $\hat{A}_m$, as in \cite{OKeeffe2018e}, such that (\ref{eqn:distance}) is satisfied. Find $P_i$, the solution to the ARE of (\ref{eq:Vdot5}).  
                \\
                (II) Select $\omega_c$, such that (\ref{eqn:L1normCond}) is satisfied.
              
            \end{algorithm} 
            
\begin{figure}[!htb]  
                        
                          \centering
                          \includegraphics[width=12cm]{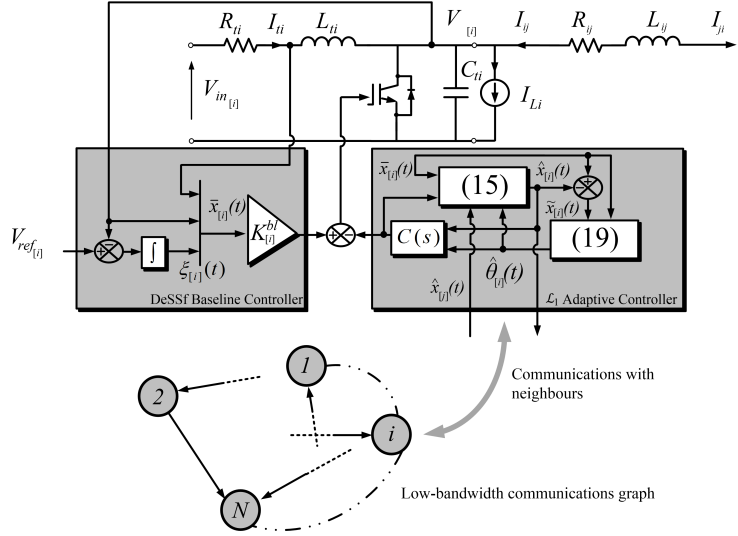}
                          \caption{Overall Distributed Control Architecture of $ \hat{\Sigma}_{[i]}^{\textrm{DGU}}$.}
                          \label{fig:OverallControlArch}
                          \end{figure}

            The overall distributed architecture is shown in Fig. \ref{fig:OverallControlArch}.

\section{Results}
A meshed and radial mG topology, similar to that of \cite{Tucci2016c}, is considered in this work. Equipped with only baseline controllers, this system is known to destabilise when $\hat{\Sigma}_{[6]}^{\textrm{DGU}}$ is plugged-in. Hence, this set-up can adequately evaluate the performance of the proposed distributed $\mathcal{L}_1$AC architecture. Each DGU is equipped with controllers $\mathcal{C}_{[i]}^{\mathcal{L}_1}, i = 1,...,6$.
\begin{figure}[!htb]    
\graphicspath{ {Images/} }
\centering
\includegraphics[width=8.2cm]{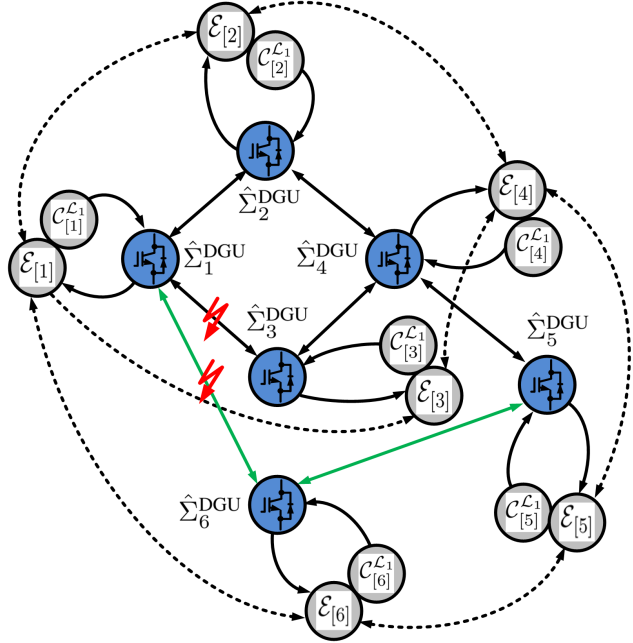}
\caption{Islanded-microgrid configuration with communications graph (dotted) - $\hat{\Sigma}_{[6]}^{\textrm{DGU}}$ plug-in (green), and topology change (red).}
\label{fig:DGU6DGU3PnPL1}
\end{figure}

Simulations are performed in Matlab/Simulink using the simpowersystems toolbox. While the averaged model of the ImG is used in \cite{OKeeffe2018e}, this paper uses the non-linear switching model for greater authenticity. System parameters are detailed in Table \ref{table:parameters}.
\begin{table}[!htb]
 \fontsize{8.5}{9.5}\selectfont
 \centering
 % used for centering table
 \caption{System Parameters} 
 \begin{tabular}{c c c c c c c c} % centered columns (4 columns)
 \hline %\hline\hline inserts double horizontal lines
  
 Description & Parameter &  $\hat{\Sigma}_{1}^{\textrm{DGU}}$ & $\hat{\Sigma}_{2}^{\textrm{DGU}}$ & $\hat{\Sigma}_{3}^{\textrm{DGU}}$ & $\hat{\Sigma}_{4}^{\textrm{DGU}}$ & $\hat{\Sigma}_{5}^{\textrm{DGU}}$ & $\hat{\Sigma}_{6}^{\textrm{DGU}}$\\ [0.5ex] % inserts table
 %heading
 \hline\hline
 
 DGU rated power (kW) & $P_{[i]}$ & 5 & 5 & 5 & 5 & 5 & 5 \\
 Local load demand (kW) & $P_{R_{[i]}}$ & 2.5 & 2 & 1.8 & 2.5 & 3 & 2.5 \\
 Input voltage (V)& $V_{in_{[i]}}$  & 95  & 100 & 90 & 105 & 92 & 90 \\
 Reference voltage (V)& $V_{ref_{[i]}}$ & 381 & 380.5 & 380.2 & 379 & 379.5 & 380.7\\
 Switching frequency (kHz)& $f_s$ & 25 & 25 & 25 & 25 & 25 & 25\\
 Duty cycle & $ D_i $ & 0.7507 & 0.7372 & 0.7633 & 0.723 & 0.7576 & 0.7636 \\
 Inductance ($\mu $H)& $L_{ti}$ & 28.47 & 89.62 & 192.5 & 70 & 35 & 93.34 \\
 Capacitance ($\mu $F) & $C_{ti}$ &  37.632 & 51.67 & 40.73 & 37 & 31 & 24.66\\
 Parasitic resistance ($\Omega$)& $R_{ti}$ & 0.02 & 0.04 & 0.02 & 0.2 & 0.4 & 0.5 \\
 Line resistance ($\Omega$)& $R_{ij}$ & 0.5-2-10 & 0.5-4 & 2-4 & 2-4-15 & 15-4 & 10-4 \\
 Line inductance ($\mu $H)& $L_{ij}$ & 10-70-800 & 40-70 & 70-70 & 70-70-25 & 25-90 & 800-90\\
 Nominal duty cycle & $ D_i $ & 0.7368
  & 0.7368
   & 0.7368
    & 0.723 & 0.7368
     & 0.7368
      \\
 Nominal inductance ($\mu $H)& $L_{t^{nom}}$ & 2.794 & 2.794 & 2.794 & 2.794 & 2.794 & 2.794 \\
 Nominal capacitance ($\mu $F) & $C_{t^{nom}}$ &  60.6 & 60.6 & 60.6 & 60.6 & 60.6 & 60.6\\
 Nominal parasitic resistance ($\Omega$)& $R_{t^{nom}}$ & 0.1 & 0.1 & 0.1 & 0.1 & 0.1 & 0.1 \\
 Nominal line resistance ($\Omega$)& $R_{ij^{nom}}$ & 1 & 1 & 1 & 1 & 1 & 1 \\
 Nominal line inductance ($\mu$H)& $L_{ij^{nom}}$ & 10 & 10 & 10 & 10 & 10 & 10 \\
 \hline
 \end{tabular}
 \label{table:parameters} % is used to refer this table in the text
 \end{table}
% Add gamma and omega_c values%

The dynamics of each DGU are different i.e. the electrical parameters and controller bandwidths are non-identical; therefore, the system can be defined as heterogeneous. The tests include PnP operations, robustness to topology change and unknown load disturbances, and output voltage reference tracking. The architecture is also evaluated using a bus-connected topology.

\subsection{Plug-and-play operations}
\begin{figure}[!htb]
\centering
{\includegraphics[width=8.5cm]{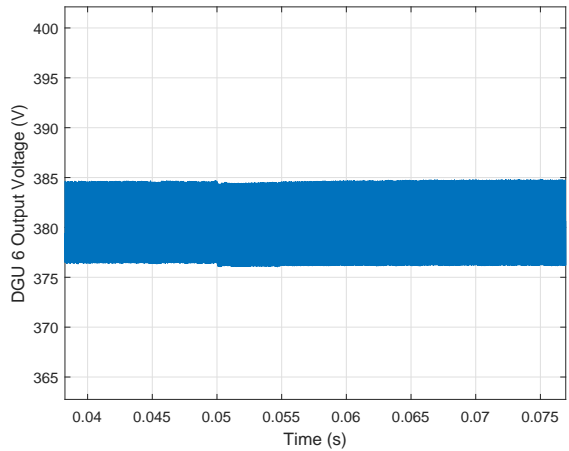}} 
\caption{$\hat{\Sigma}_{[6]}^{\textrm{DGU}}$ voltage response during plug-in.}
\label{fig:DGUPnP}
\end{figure}

At $t = 0.05$ s, $\hat{\Sigma}_{[6]}^{\textrm{DGU}}$ is plugged-in to the network. Fig. \ref{fig:DGUPnP}(a) shows that $\hat{\Sigma}_{[6]}^{\textrm{DGU}}$ seamlessly plugs-in and remains stable thereafter.

\begin{figure}[!htb]
\centering
{\includegraphics[width=8.5cm]{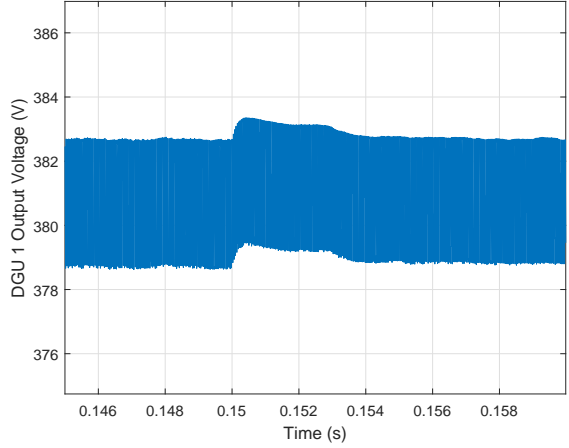}} 
\caption{ $\hat{\Sigma}_{[1]}^{\textrm{DGU}}$during topology change.}
\label{fig:DGUFault}
\end{figure}

At $t=0.15$ s, lines connecting $\hat{\Sigma}_{[1]}^{\textrm{DGU}}$ to $\hat{\Sigma}_{[3]}^{\textrm{DGU}}$ and $\hat{\Sigma}_{[6]}^{\textrm{DGU}}$ are disconnected due to a fault. Thus, the topology of the ImG changes; it is no longer radial. Fig. \ref{fig:DGUFault} shows $\hat{\Sigma}_{[1]}^{\textrm{DGU}}$ has fast and robust behaviour;
after a 1 V overshoot, the response settles within 5 ms.

\subsection{Robustness to unknown load changes}
\begin{figure}[!htb]
\centering
\includegraphics[width=8.5cm]{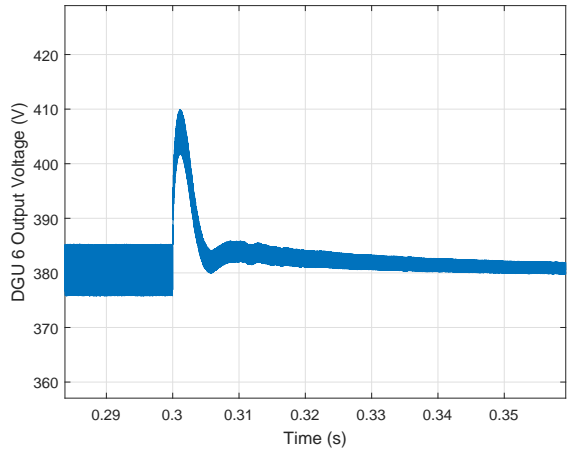}
\caption{$\hat{\Sigma}_{[6]}^{\textrm{DGU}}$ voltage response for 2.5 kW - 800 W load change.}
\label{DGU6Load}
\end{figure}

Robustness to an unknown load change is evaluated by stepping the load power of $\hat{\Sigma}_{[6]}^{\textrm{DGU}}$ at $t = 0.3$ s from 2.5 kW to 800 W. Fig. \ref{DGU6Load} shows that though overshoot amounts to 7.8 \%, settling is achieved within 30 ms.

\subsection{Output voltage reference tracking}

The hierarchical structure of mG control structures requires primary level reference changes. Commands are sent from secondary controllers, in order to control the power flows amongst DGUs within the mG, as well as regulate the state-of-charge of batteries. Therefore, a key metric of the proposed system is the performance of the system in response to voltage reference changes. This is evaluated by stepping the voltage reference of $\hat{\Sigma}_{[5]}^{\textrm{DGU}}$ from 379.5 V to 377 V.

\begin{figure}[!htb]
\centering
\includegraphics[width=8.5cm]{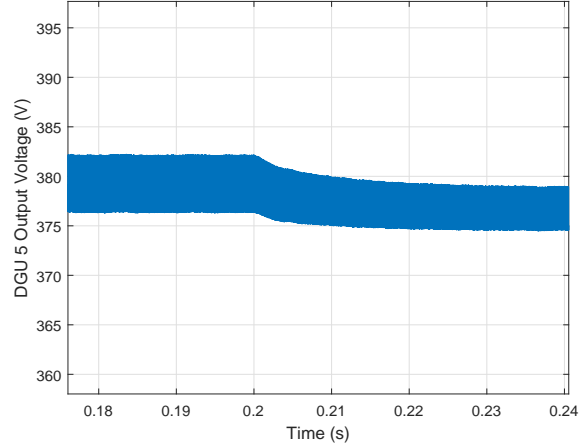}
\caption{$\hat{\Sigma}_{[5]}^{\textrm{DGU}}$ voltage reference step change of 379 V - 377 V.}
\label{DGU6Vref}
\end{figure}

\subsection{Bus-connected topology}
\begin{figure}[!htb]
\centering
\includegraphics[width=10cm]{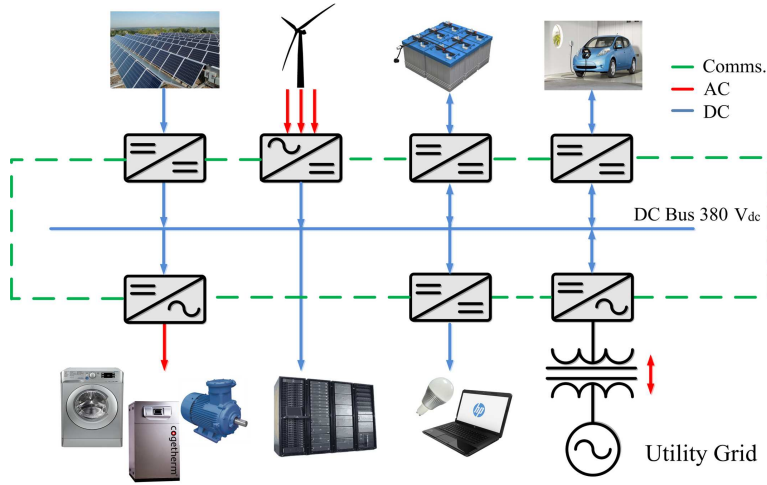}
\caption{Typical bus-connected topology, adapted from \cite{Lu2014}.}
\label{fig:BusConnect}
\end{figure}

 A bus-connected topology is used to highlight the flexibility of designing controls for arbitrary topologies. 
 \vspace{3mm} \newline
 \textbf{Remark 2.} \textit{Since typical mG loads are not directly connected to each DGU i.e. bus-connected topology, Kron reduction analysis is required to transpose the coupling parameters of the bus-connected into those of the load-connected topology of Fig.  \ref{fig:MG2} that the design \label{eqn:ctrl} is based on. The advantage of the proposed adaptive design over state-of-the-art PnP techniques is that robustness to the change of parameter values can be incorporated.}
  \vspace{3mm} \newline
 Fig. \ref{fig:BusConnect} shows a typical bus-connected topology.

 Following \cite{OKeeffe2017a}, in which a resistive load of 15 kW and a 3.8 kW closed-loop speed controlled DC motor are connected to the 380 V bus, and powered by 2 DGUs, we implement the same but with 6 DGUs connected. Fig. \ref{fig:DGU6PnP}(f) shows the response of $\hat{\Sigma}_{[6]}^{\textrm{DGU}}$ plugging-in at $t = 0.1$ s. Stability is maintained during PnP operations and responses settle within 20 ms. 

\begin{figure}[!htb] % "[t!]" placement specifier just for this example
\begin{subfigure}{0.48\textwidth}
\includegraphics[width=\linewidth]{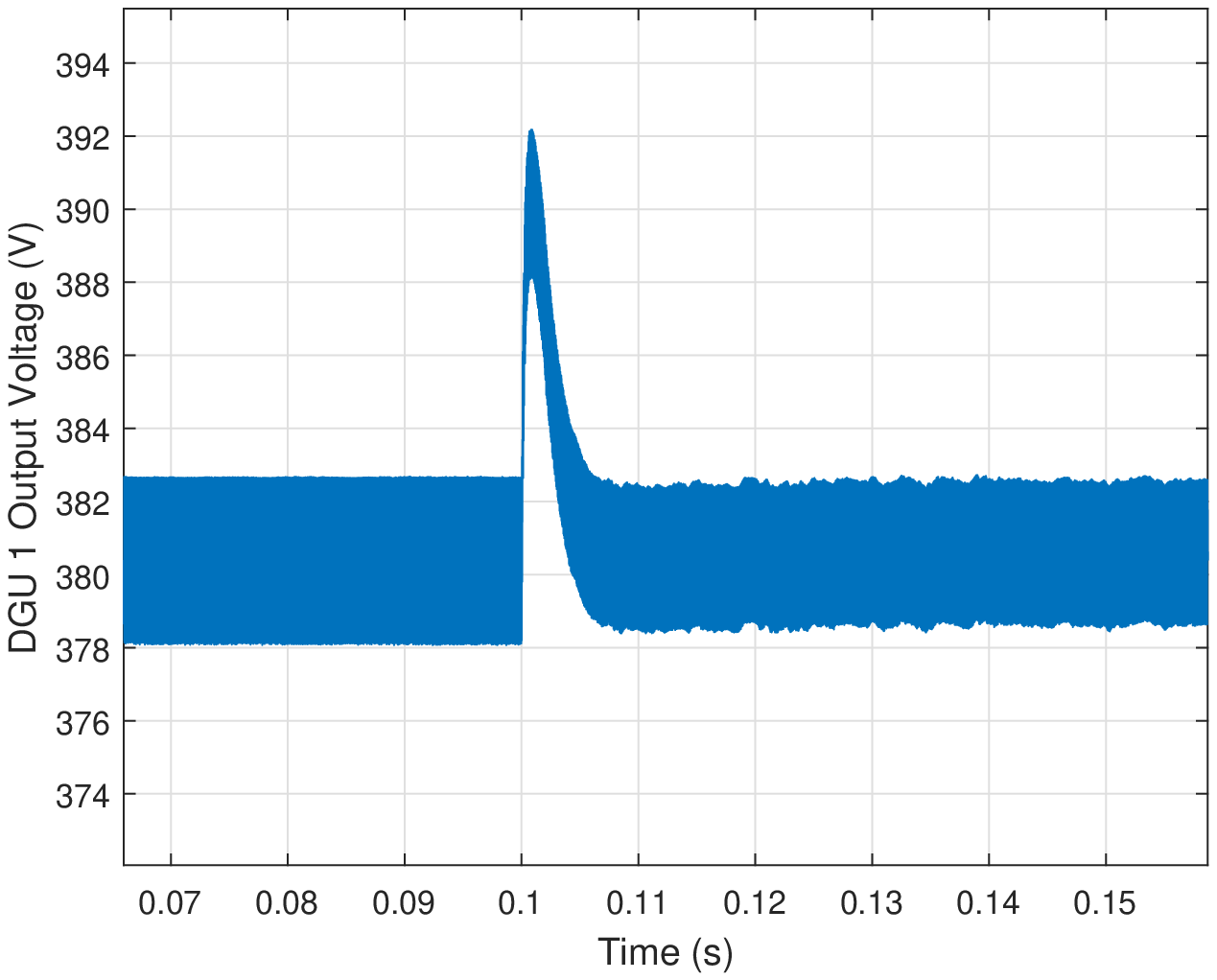}
\caption{$\hat{\Sigma}_{1}^{\textrm{DGU}}$ output voltage} \label{fig:DGU1_d}
\end{subfigure}\hspace*{\fill}
\begin{subfigure}{0.48\textwidth}
\includegraphics[width=\linewidth]{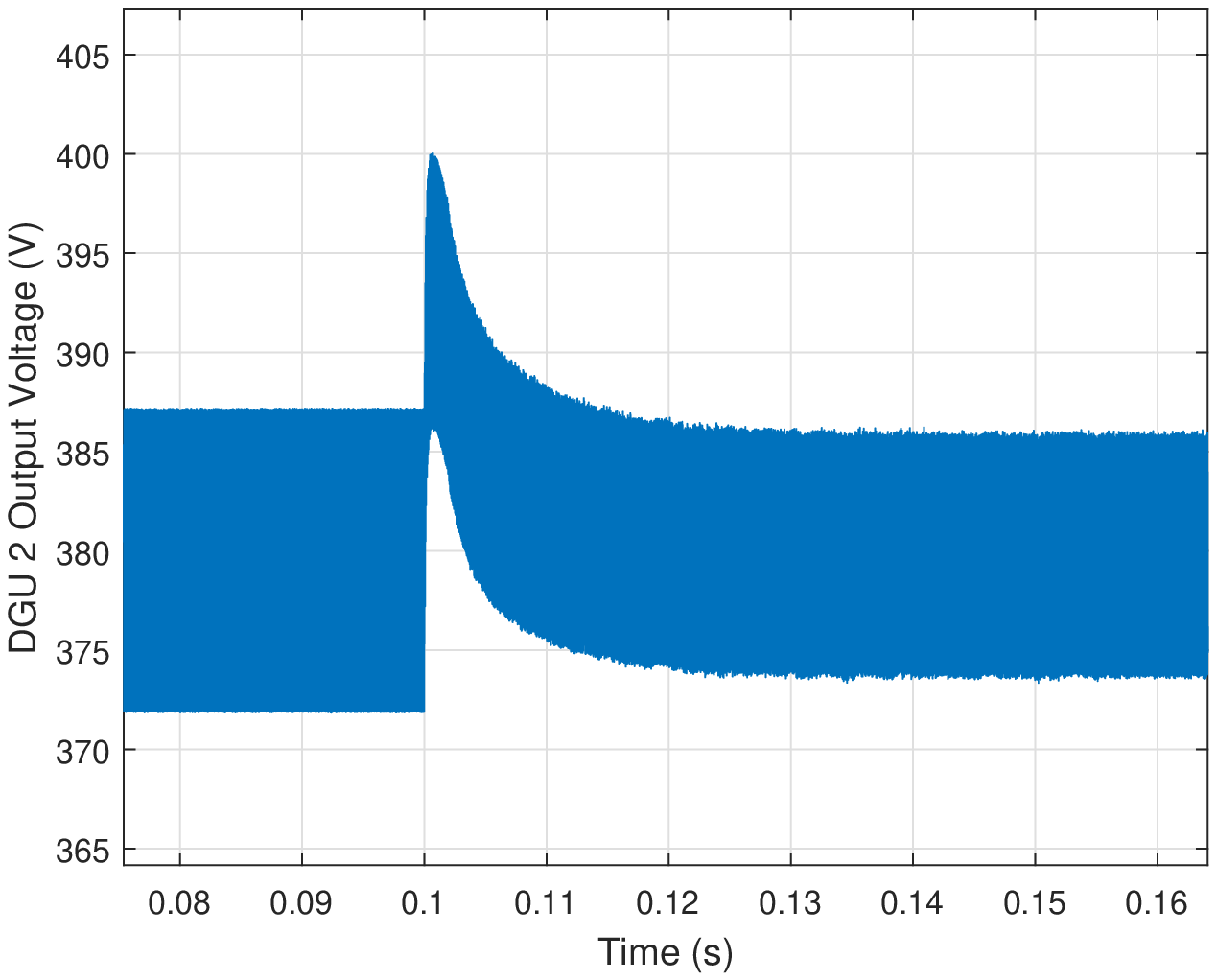}
\caption{$\hat{\Sigma}_{2}^{\textrm{DGU}}$ output voltage} \label{fig:DGU2_d}
\end{subfigure}
\medskip
\begin{subfigure}{0.48\textwidth}
\includegraphics[width=\linewidth]{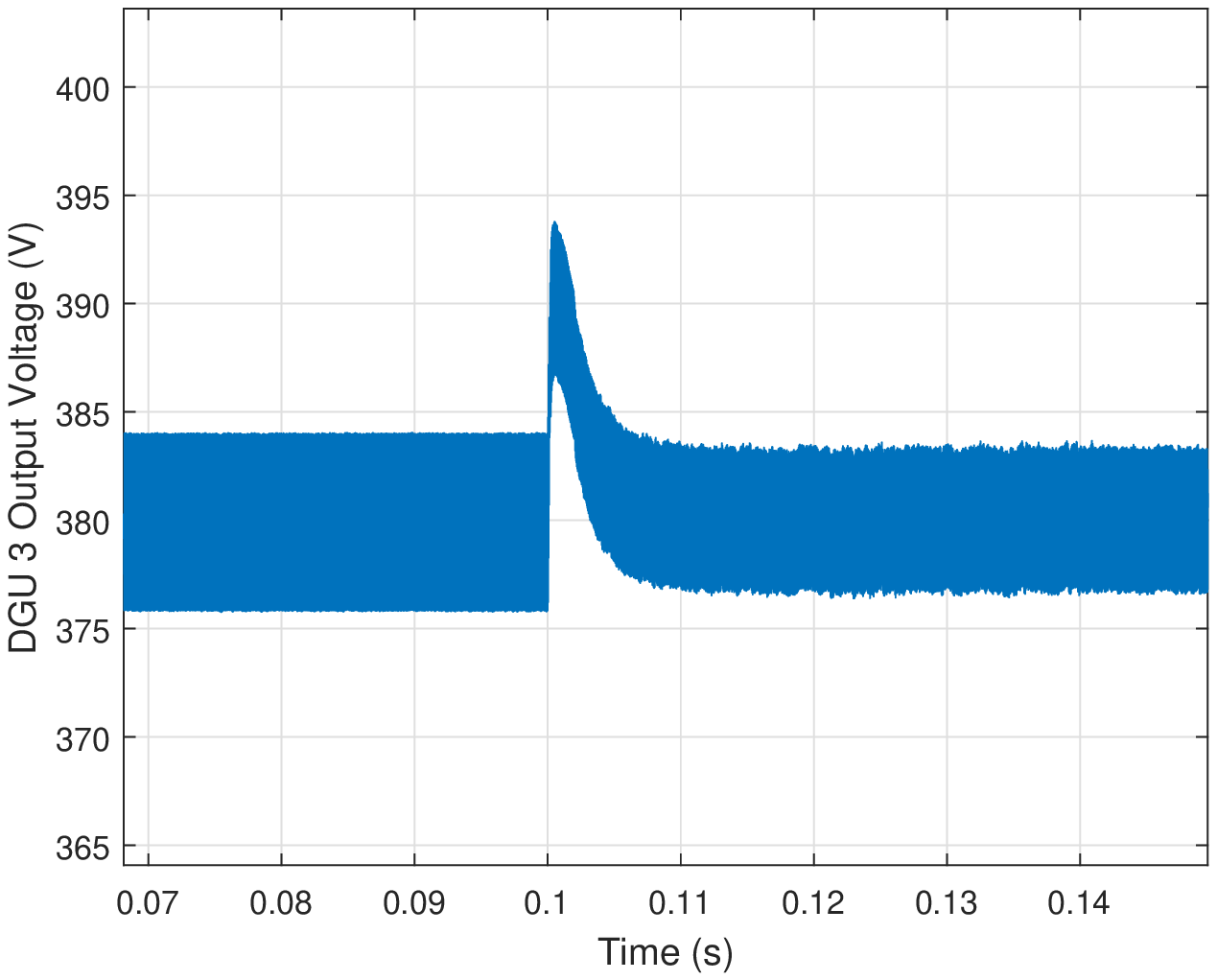}
\caption{$\hat{\Sigma}_{3}^{\textrm{DGU}}$ output voltage} \label{fig:DGU3_d}
\end{subfigure}\hspace*{\fill}
\begin{subfigure}{0.48\textwidth}
\includegraphics[width=\linewidth]{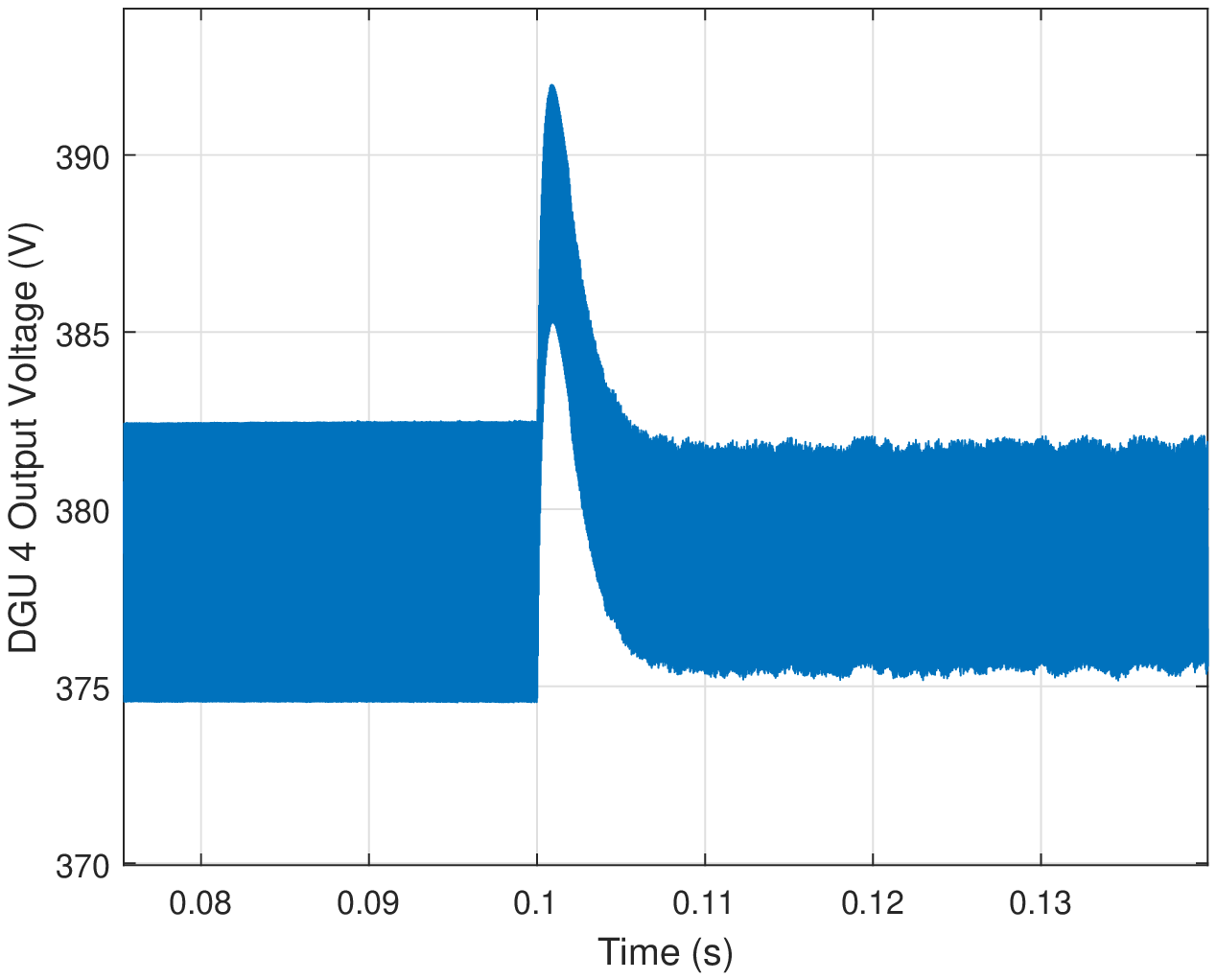}
\caption{$\hat{\Sigma}_{4}^{\textrm{DGU}}$ output voltage} \label{fig:DGU4_d}
\end{subfigure}
\end{figure}
\begin{figure}[!htb] % "[t!]" placement specifier just for this example
\ContinuedFloat%
\graphicspath{ {Images/} }
\begin{subfigure}{0.48\textwidth}
\includegraphics[width=\linewidth]{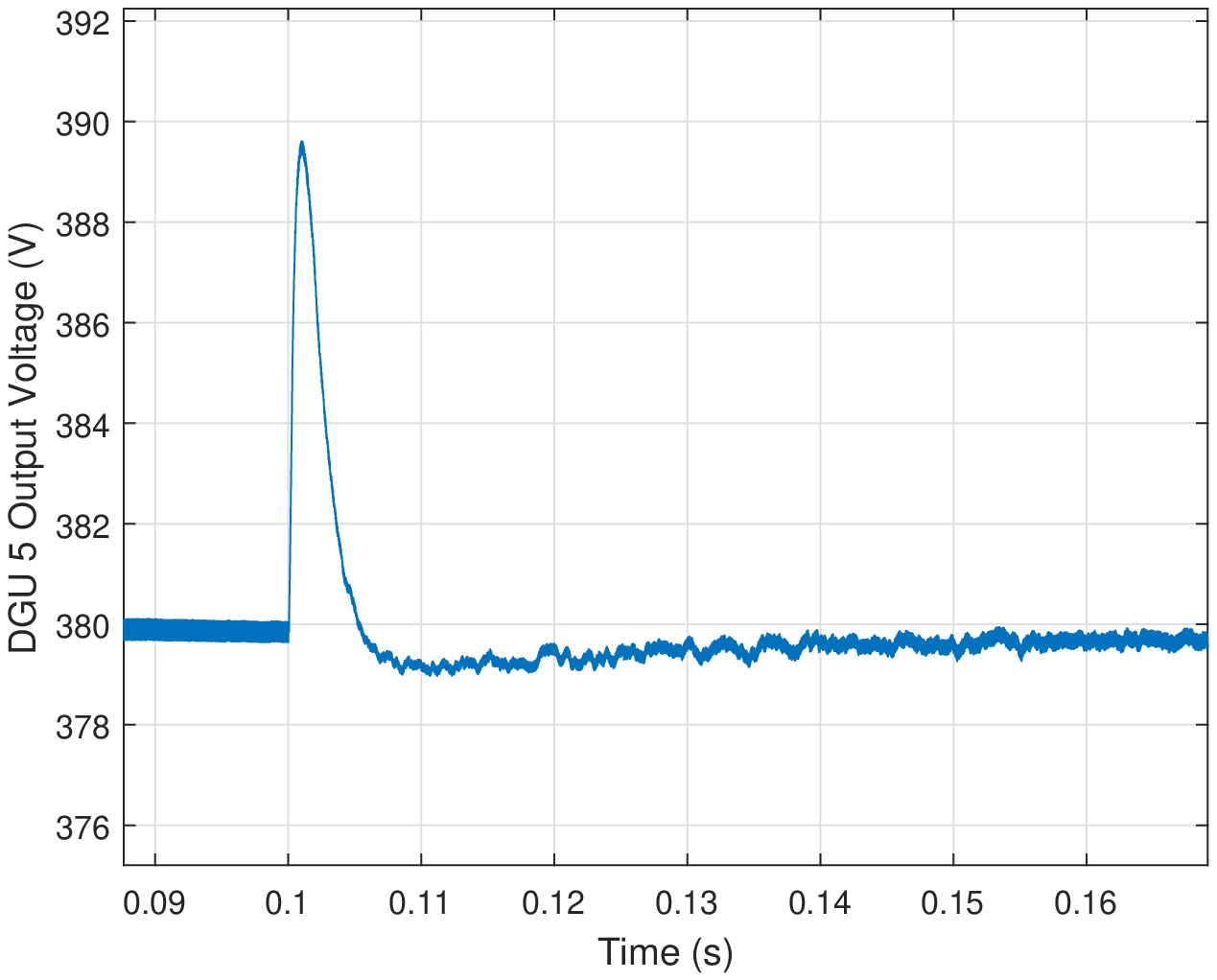}
\caption{$\hat{\Sigma}_{5}^{\textrm{DGU}}$ output voltage} \label{fig:DGU5_d}
\end{subfigure}\hspace*{\fill}
\begin{subfigure}{0.48\textwidth}
\includegraphics[width=\linewidth]{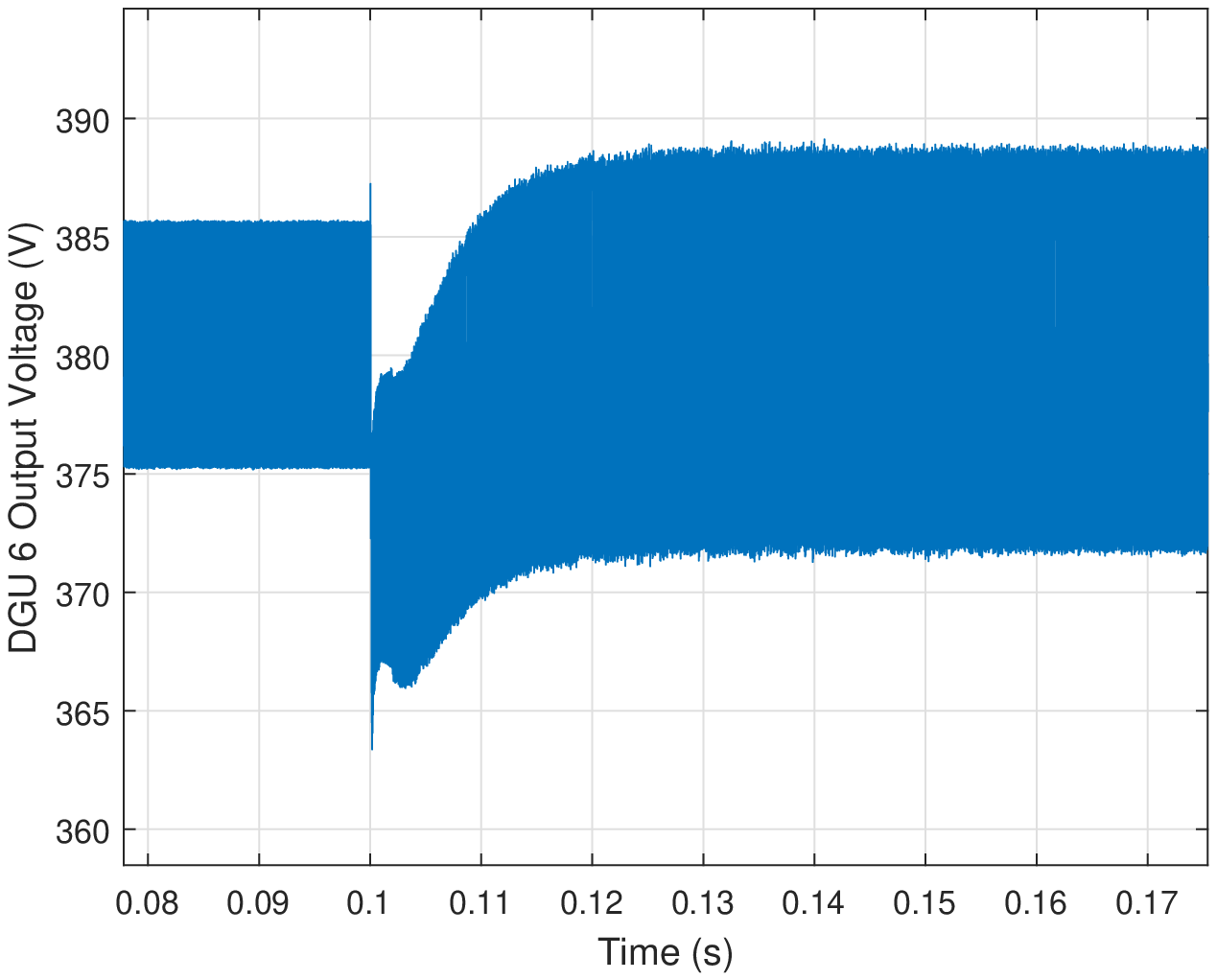}
\caption{$\hat{\Sigma}_{6}^{\textrm{DGU}}$ output voltage} \label{fig:DGU6_d}
\end{subfigure}
\caption{DGU output voltage responses in bus-connected topology to $\hat{\Sigma}_{[6]}^{\textrm{DGU}}$ plug-in.} \label{fig:DGU6PnP}
\end{figure}

\vspace{10mm}
At $t = 0.2$ s, $\hat{\Sigma}_{[3]}^{\textrm{DGU}}$ is plugged-out. Fig. \ref{fig:DGU3PnP} plots the responses of all the DGUs.

\begin{figure}[!htb] % "[t!]" placement specifier just for this example
\begin{subfigure}{0.48\textwidth}
\includegraphics[width=\linewidth]{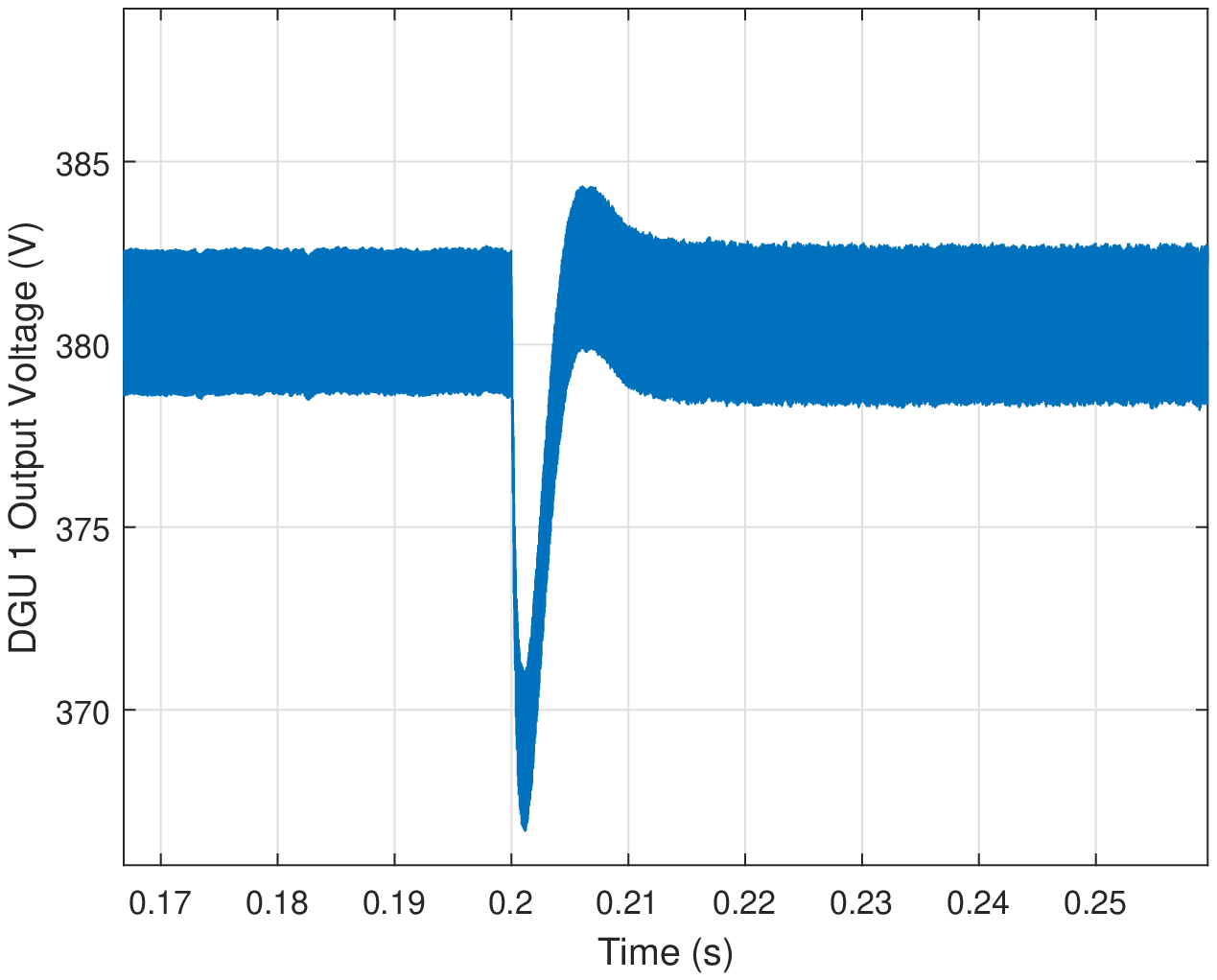}
\caption{$\hat{\Sigma}_{1}^{\textrm{DGU}}$ output voltage} \label{fig:DGU1_d}
\end{subfigure}\hspace*{\fill}
\begin{subfigure}{0.48\textwidth}
\includegraphics[width=\linewidth]{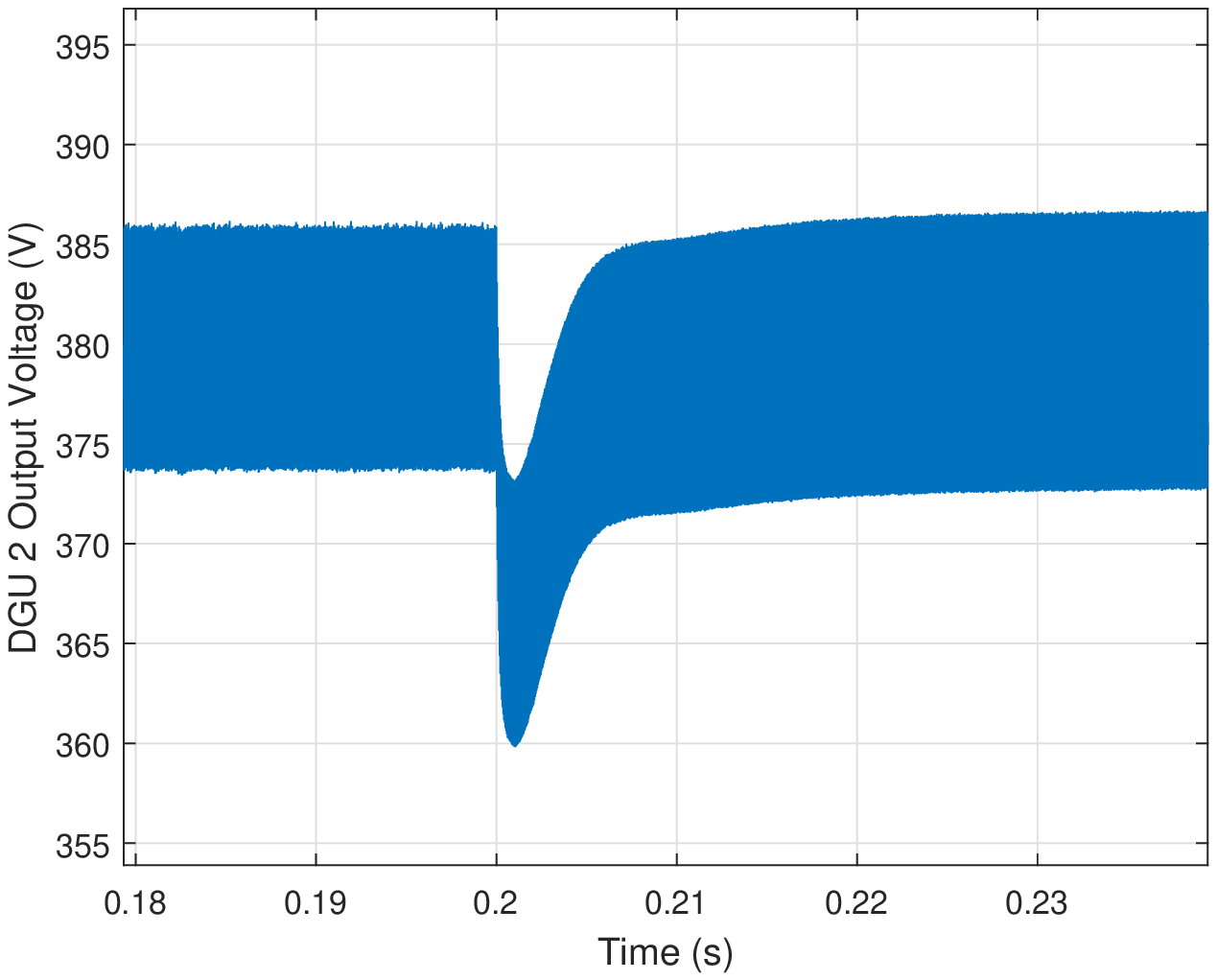}
\caption{$\hat{\Sigma}_{2}^{\textrm{DGU}}$ output voltage} \label{fig:DGU2_d}
\end{subfigure}
\medskip
\begin{subfigure}{0.48\textwidth}
\includegraphics[width=\linewidth]{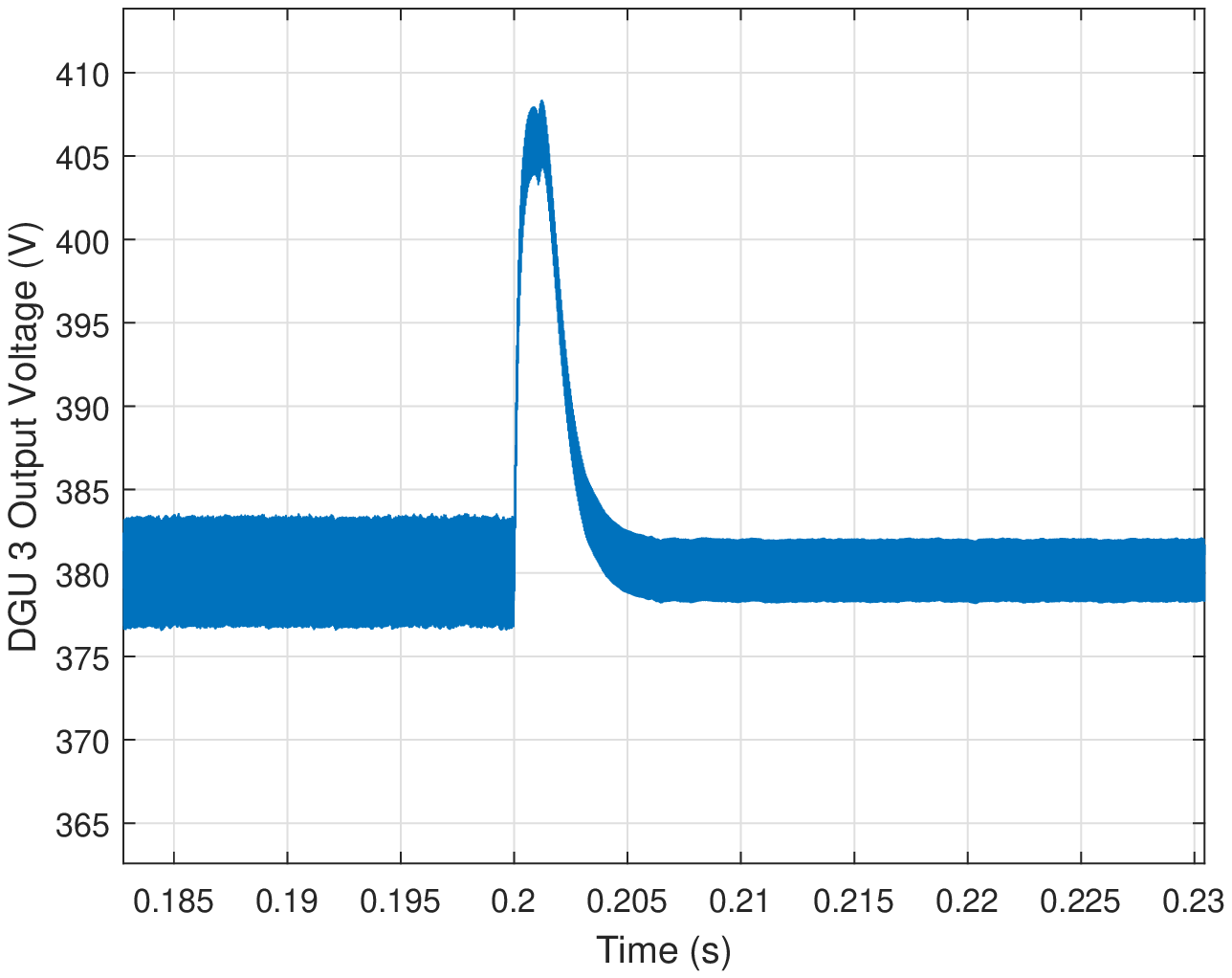}
\caption{$\hat{\Sigma}_{3}^{\textrm{DGU}}$ output voltage} \label{fig:DGU3_d}
\end{subfigure}\hspace*{\fill}
\begin{subfigure}{0.48\textwidth}
\includegraphics[width=\linewidth]{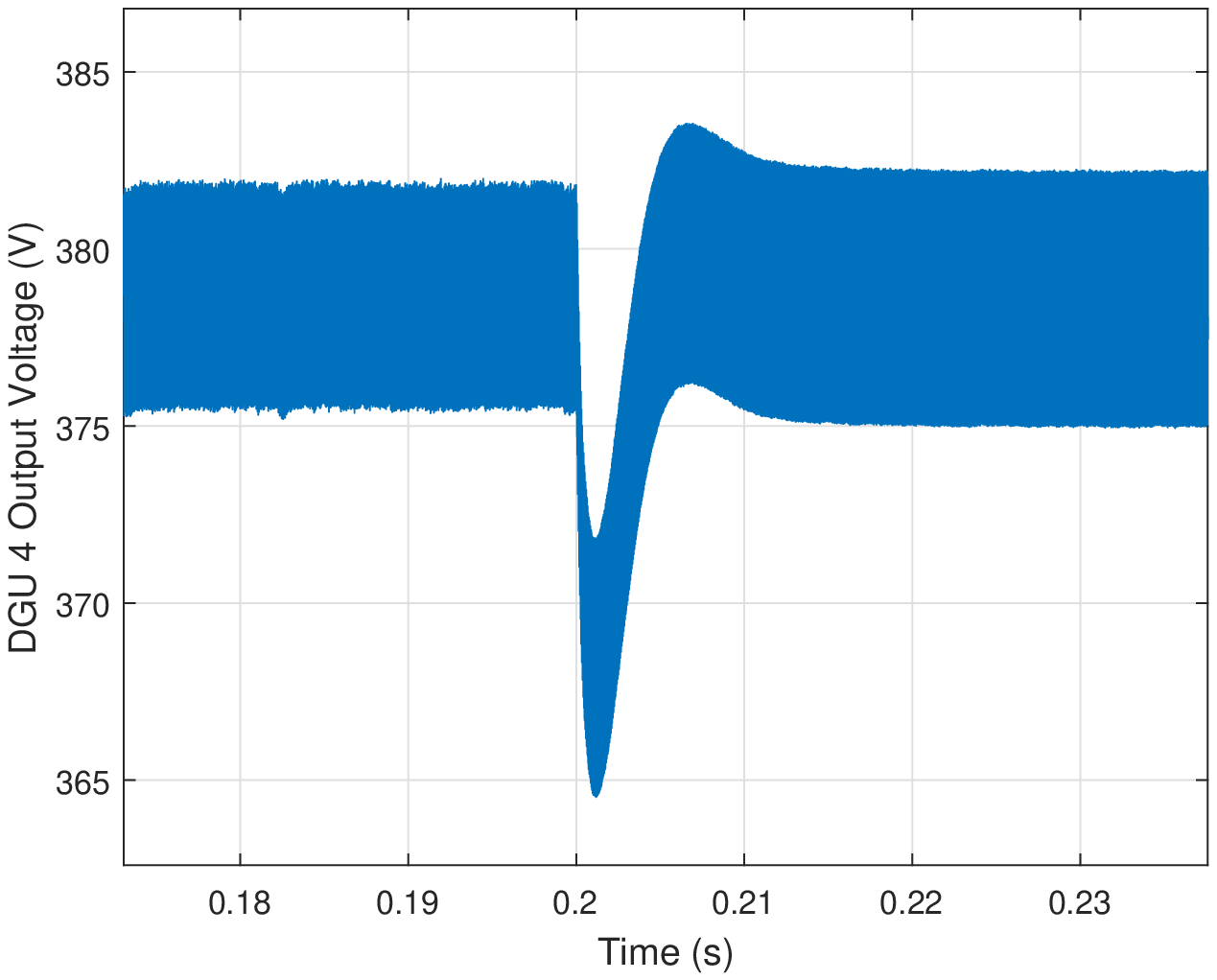}
\caption{$\hat{\Sigma}_{4}^{\textrm{DGU}}$ output voltage} \label{fig:DGU4_d}
\end{subfigure}
\end{figure}
\begin{figure}[!htb] % "[t!]" placement specifier just for this example
\ContinuedFloat%
\graphicspath{ {Images/} }
\begin{subfigure}{0.48\textwidth}
\includegraphics[width=\linewidth]{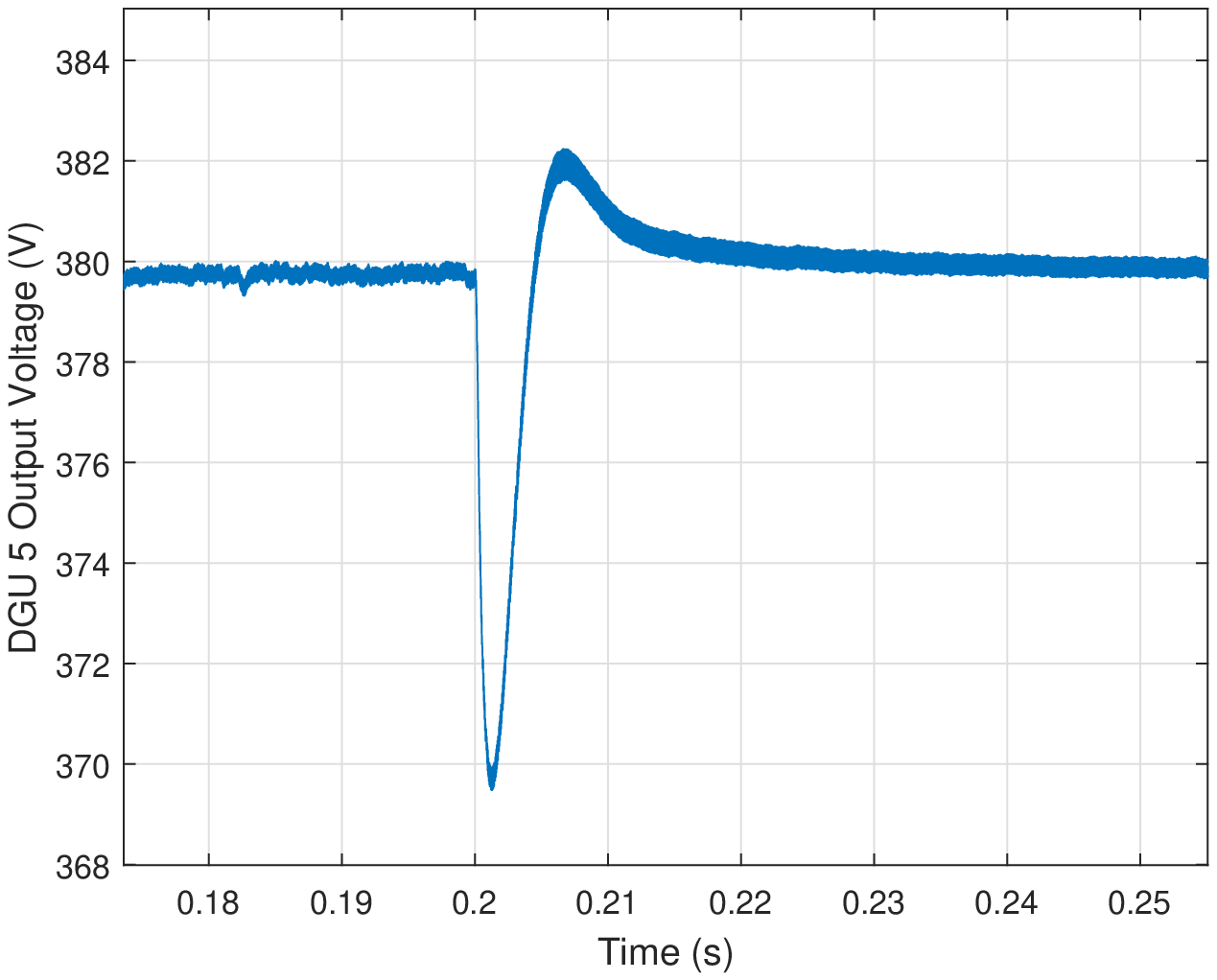}
\caption{$\hat{\Sigma}_{5}^{\textrm{DGU}}$ output voltage} \label{fig:DGU5_d}
\end{subfigure}\hspace*{\fill}
\begin{subfigure}{0.48\textwidth}
\includegraphics[width=\linewidth]{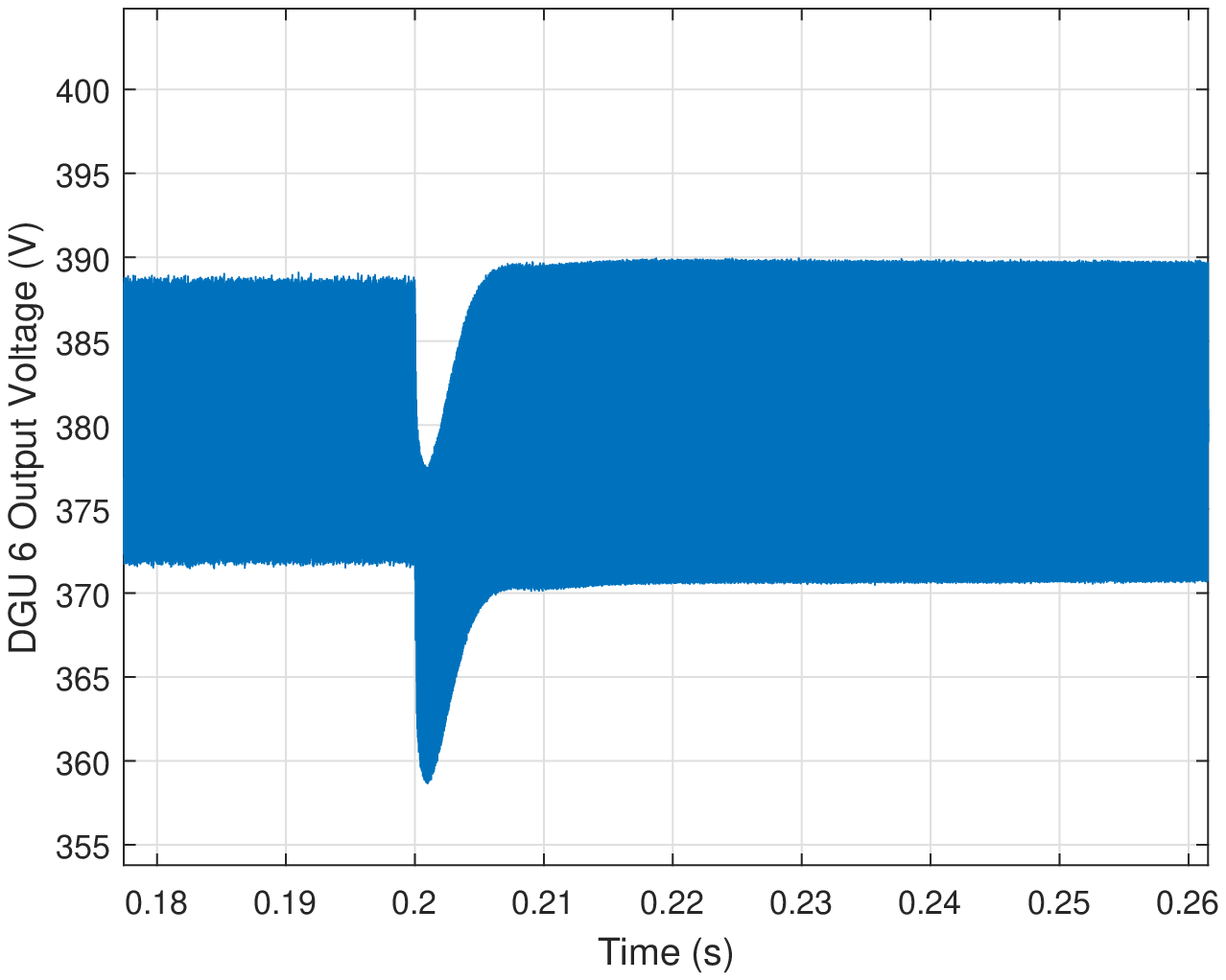}
\caption{$\hat{\Sigma}_{6}^{\textrm{DGU}}$ output voltage} \label{fig:DGU6_d}
\end{subfigure}
\caption{DGU output voltage responses to $\hat{\Sigma}_{[3]}^{\textrm{DGU}}$ plug-out in bus-connected topology.} \label{fig:DGU3PnP}
\end{figure}

Again, stability is maintained, with the worst case response associated with $\hat{\Sigma}_{5}^{\textrm{DGU}}$ in Fig. \ref{fig:DGU3PnP}(e) which shows a settling time of nearly 30 ms and $\hat{\Sigma}_{6}^{\textrm{DGU}}$ in Fig. \ref{fig:DGU3PnP}(f).

\vspace{50mm}
Finally, the resistive load power is changed from 15 kW to 18 kW.
\newpage
\begin{figure}[!htb] % "[t!]" placement specifier just for this example
\begin{subfigure}{0.48\textwidth}
\includegraphics[width=\linewidth]{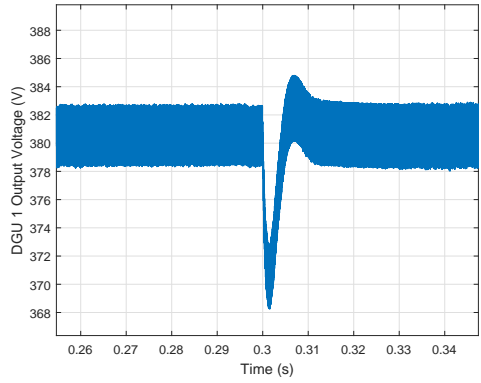}
\caption{$\hat{\Sigma}_{[1]}^{\textrm{DGU}}$ output voltage.} \label{fig:DGU1_d}
\end{subfigure}\hspace*{\fill}
\begin{subfigure}{0.48\textwidth}
\includegraphics[width=\linewidth]{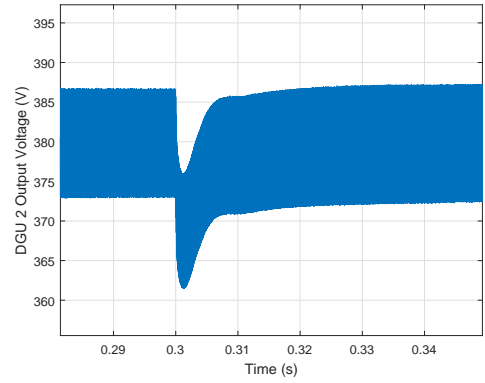}
\caption{$\hat{\Sigma}_{[2]}^{\textrm{DGU}}$ output voltage.} \label{fig:DGU2_d}
\end{subfigure}
\medskip
\begin{subfigure}{0.48\textwidth}
\includegraphics[width=\linewidth]{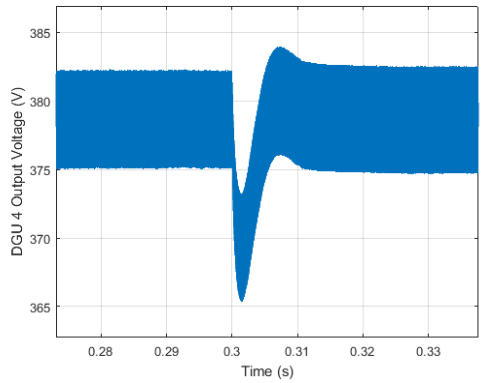}
\caption{$\hat{\Sigma}_{[4]}^{\textrm{DGU}}$ output voltage.} \label{fig:DGU3_d}
\end{subfigure}\hspace*{\fill}
\begin{subfigure}{0.48\textwidth}
\includegraphics[width=\linewidth]{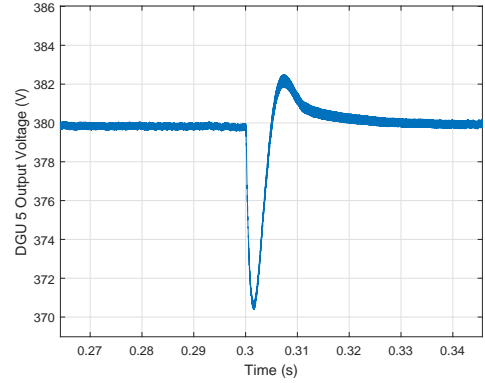}
\caption{$\hat{\Sigma}_{[5]}^{\textrm{DGU}}$ output voltage.} \label{fig:DGU4_d}
\end{subfigure}
\medskip
\centering
\begin{subfigure}{0.48\textwidth}
\includegraphics[width=\linewidth]{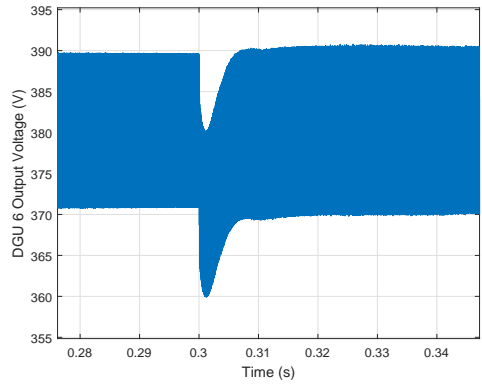}
\caption{$\hat{\Sigma}_{[6]}^{\textrm{DGU}}$ output voltage} \label{fig:DGU5_d}
\end{subfigure}
\caption{DGU output voltage responses to 15 kW - 18 kW load step change in bus-connected topology.} \label{fig:LoadChange}
\end{figure}

Fig. \ref{fig:LoadChange}(e) shows the response of $\hat{\Sigma}_{[6]}^{\textrm{DGU}}$ settling within 15 ms.

\newpage

\section{Conclusion}
This paper develops a novel scalable distributed $\mathcal{L}_{1}$AC architecture for the primary voltage control level of large-scale DC ImGs with arbitrary topology. $\mathcal{L}_1$ adaptive controllers were designed with self-commissioning capabilities  in that existing baseline controllers can be augmented and GAS is guaranteed in a plug-and-play fashion. The distributed architecture is shown to ensure GAS in the presence of large-gain interconnections and parametric, topology, and PnP operation uncertainty. This requires knowledge of the upper-bound on interconnection terms and measurement of neighbouring predictor states in order to solve local AREs. 

The architecture was successfully evaluated using a heterogeneous radial and meshed ImG that consisted of DC-DC boost converters. Fast and robust voltage control is achieved during PnP operations, topology changes and unknown load disturbances. The flexibility of the control orientated design approach was evaluated using a bus-connected topology consisting of resistive and closed-loop controlled DC motor loads. Kron reduction analysis, typically required by state-of-the-art PnP techniques, can be relaxed due to the adaptive nature of the controller. Ultimately, with the presence of LBC in the conventional secondary and tertiary levels of the mG control hierarchy, a distributed primary control architecture is feasible.  

Future work will continue to investigate cases of robustness as further conditions are required for coupling-independence, and for the architecture to provide stability in mGs with constant-power loads. Moreover, the proposed architecture will be designed with purely adaptive controllers, implemented for current controls and evaluated when equipped with secondary/coordination control layers.

\bibliography{Smart_Grid_Paper}
\end{document}